\documentclass[a4paper]{aa}
\usepackage{epstopdf}
\usepackage{amsmath}
\usepackage{amsfonts}
\usepackage{amssymb}
\usepackage{graphicx}
\usepackage{txfonts}
\usepackage{natbib}
\usepackage{siunitx}
\bibpunct{(}{)}{;}{a}{}{,} 

\DeclareSIUnit\parsec{pc}

\begin{document}
\DeclareGraphicsExtensions{.pdf,.gif,.jpg}
 \bibliographystyle{aa}

\title{A weak lensing analysis of the PLCK G100.2-30.4 cluster. 
\thanks{Based  on data collected at Subaru Telescope  (University of Tokyo).}
}
\author{M. Radovich \inst{1} \and I. Formicola\inst{2} 
\and M. Meneghetti\inst{2}
\and I. Bartalucci\inst{3,7}
\and H. Bourdin  \inst{3}
\and P. Mazzotta \inst{3}
\and L. Moscardini \inst{9,2,10}
\and S. Ettori \inst{2,10}
\and M. Arnaud \inst{7}
\and G. W. Pratt\inst{7}
\and N. Aghanim \inst{5}
\and H. Dahle \inst{4}
\and M. Douspis \inst{5}
\and E. Pointecouteau \inst{6,11}
\and A. Grado \inst{8} }

\institute{
 INAF - Osservatorio Astronomico di Padova, vicolo dell'Osservatorio 5, I-35122, Padova, Italy
\and INAF - Osservatorio Astronomico di Bologna, via Ranzani 1, I-40127, Bologna, Italy 
\and Dipartimento di Fisica, Universit\`{a} Tor Vergata, via della Ricerca Scientifica 1, I-00133 Roma, Italy
\and  Institute of Theoretical Astrophysics, University of Oslo, Blindern, Oslo, Norway 
\and Institut d'Astrophysique Spatiale, CNRS (UMR8617) Universit\'{e} Paris-Sud 11, B\^{a}t. 121, F-91405 Orsay, France 
\and CNRS, IRAP, 9 avenue colonel Roche, BP 44346, F-31028 Toulouse cedex 4, France
\and  Laboratoire AIM, IRFU/Service d'Astrophysique, CEA/DSM, CNRS, Universit\'{e} Paris Diderot, B\^{a}t. 709, CEA-Saclay, F-91191 Gif-sur-Yvette Cedex, France 
\and INAF - Osservatorio Astronomico di Capodimonte, Salita Moiariello 16, I-80131, Napoli
\and  Dipartimento di Fisica e Astronomia Alma Mater Studiorum - Universit\'{a} di Bologna, viale Berti Pichat 6/2, I-40127 Bologna, Italy
\and INFN - Sezione di Bologna, viale Berti-Pichat 6/2, I-40127 Bologna, Italy 
\and  Universit\'{e} de Toulouse, UPS-OMP, IRAP, F-31028 Toulouse cedex 4, France }
\date{received; accepted}

 \abstract
   {   
 We present a mass estimate of the Planck-discovered cluster PLCK G100.2-30.4, derived from a weak  lensing analysis of deep SUBARU $griz$ images. We perform a  careful selection of the background galaxies using the multi-band imaging data, and undertake the  weak lensing analysis on the deep (1hr) $r-$band image. The shape measurement is based on the KSB algorithm; we adopt the PSFex software to model the Point Spread Function (PSF) across the field and correct for this in the shape measurement. The weak lensing analysis is validated through extensive image simulations. We compare the resulting weak lensing mass profile and total mass estimate to those obtained from our re-analysis of XMM-Newton observations, derived under the hypothesis of hydrostatic equilibrium.  The total integrated mass profiles are in remarkably good agreement, agreeing within $1\sigma$ across their common radial range. A mass $M_{\rm 500} \sim 7 \times 10^{14}$ $M_\odot$  is derived for the cluster from our weak lensing analysis.  Comparing this value to that obtained from our reanalysis of XMM-Newton data, we obtain a bias factor of (1-$b$) = 0.8 $\pm$ 0.1.
 This is compatible within $1 \sigma$ with the value of (1-$b$) obtained by  \citeauthor{planck2015-p24} from their calibration of the bias factor using newly-available weak lensing reconstructed masses.}
    

\maketitle

\section{Introduction}
\label{sec:intro}

Clusters of galaxies are a crucial probe for a variety of key cosmological 
issues. Among their many  applications, knowledge of the mass function and  its redshift evolution allows constraints to be put on the matter power spectrum and limits to be set on 
several cosmological parameters. The construction of the cluster mass function  generally involves the following steps: 1) clusters should be identified in large
surveys; 2) mass proxies should be used to estimate their masses; and 3) the
mass function should be constructed by counting clusters in given mass and
redshift bins. This process relies on the fundamental assumption that the
scaling relations between the mass and its proxies are well-calibrated and
unaffected by selection biases (or that the biases have been accounted for). It has been successfully implemented both using X-rays surveys as the ROSAT All-Sky Survey, where
proxies like the gas mass, the temperature, or the X-ray luminosity have been used to
estimate the masses \citep[e.g.,][]{2009ApJ...692.1060V,2010MNRAS.406.1759M}, and in optical photometric optical surveys (e.g. the SDSS), where the
masses are usually derived from the cluster richness \citep[e.g.,][]{2009ApJ...703..601R}. 
More recently, searches for galaxy clusters have been started via the detection of the
Sunyaev-Zel'dovich (SZ) effect on the photons of the Cosmic-Microwave-Background. The newly-established millimetric observatories, both ground- (South Pole Telescope and Atacama Cosmology Telescope) and space-based (Planck), have provided catalogues of hundreds to thousands of clusters. The first cluster detections from the Planck all-sky survey were released in the Planck Early SZ sample \citep{planck2011-5.1a}. A crucial part of the validation process involved use of  snapshot XMM-Newton observations; in the first two campaigns, 21 out of 25 newly identified cluster candidates were confirmed to be real \citep{planck2011-5.1b}.

\begin{figure}
 \centering
 \includegraphics[width=9.5 cm]{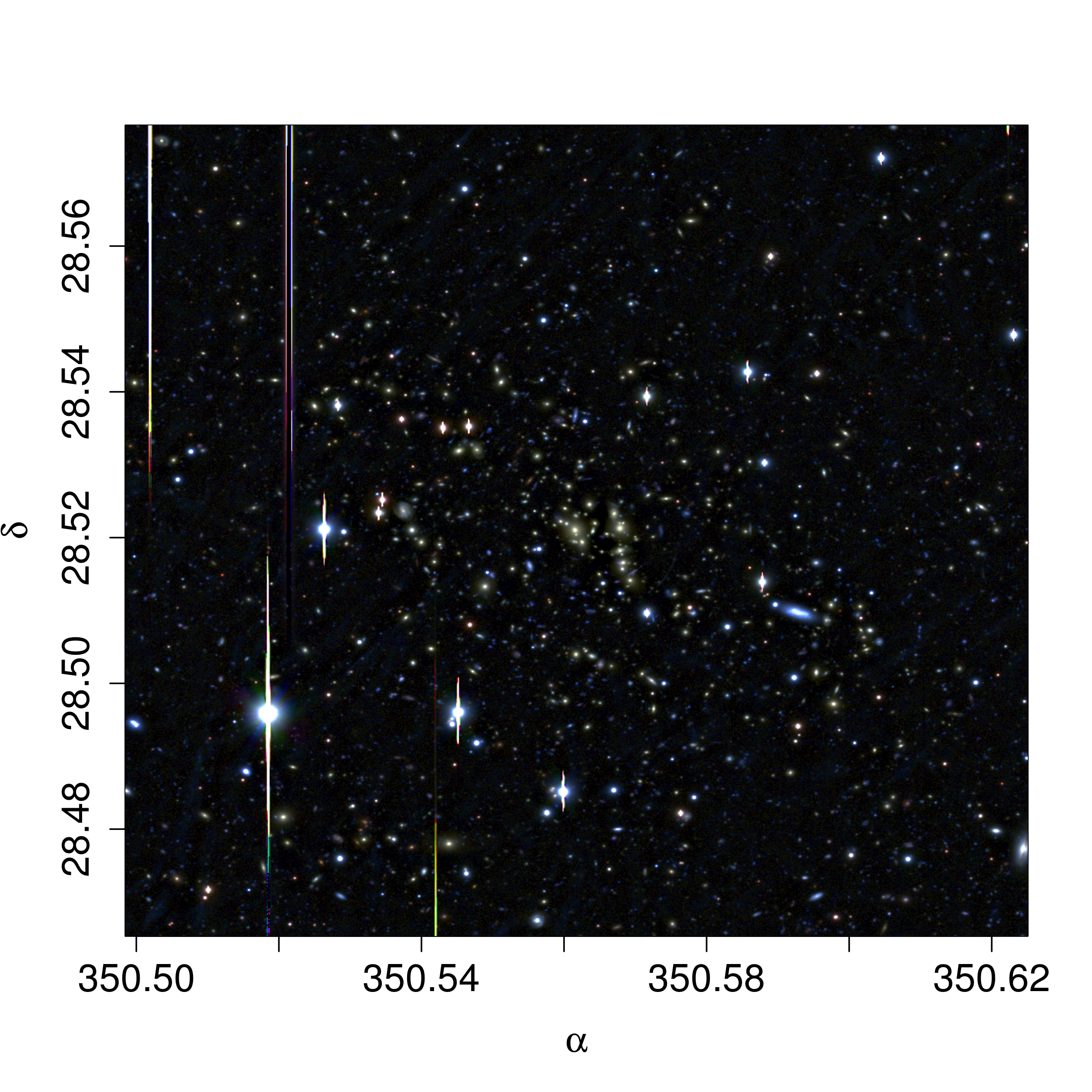}
 \caption{Composite $gri$ image of the inner $400\arcsec \times 400\arcsec$ region of the PLCKG100 field.  }
 \label{fig:field}
\end{figure}

  \begin{figure}
   \centering
   \includegraphics[width=9.5 cm]{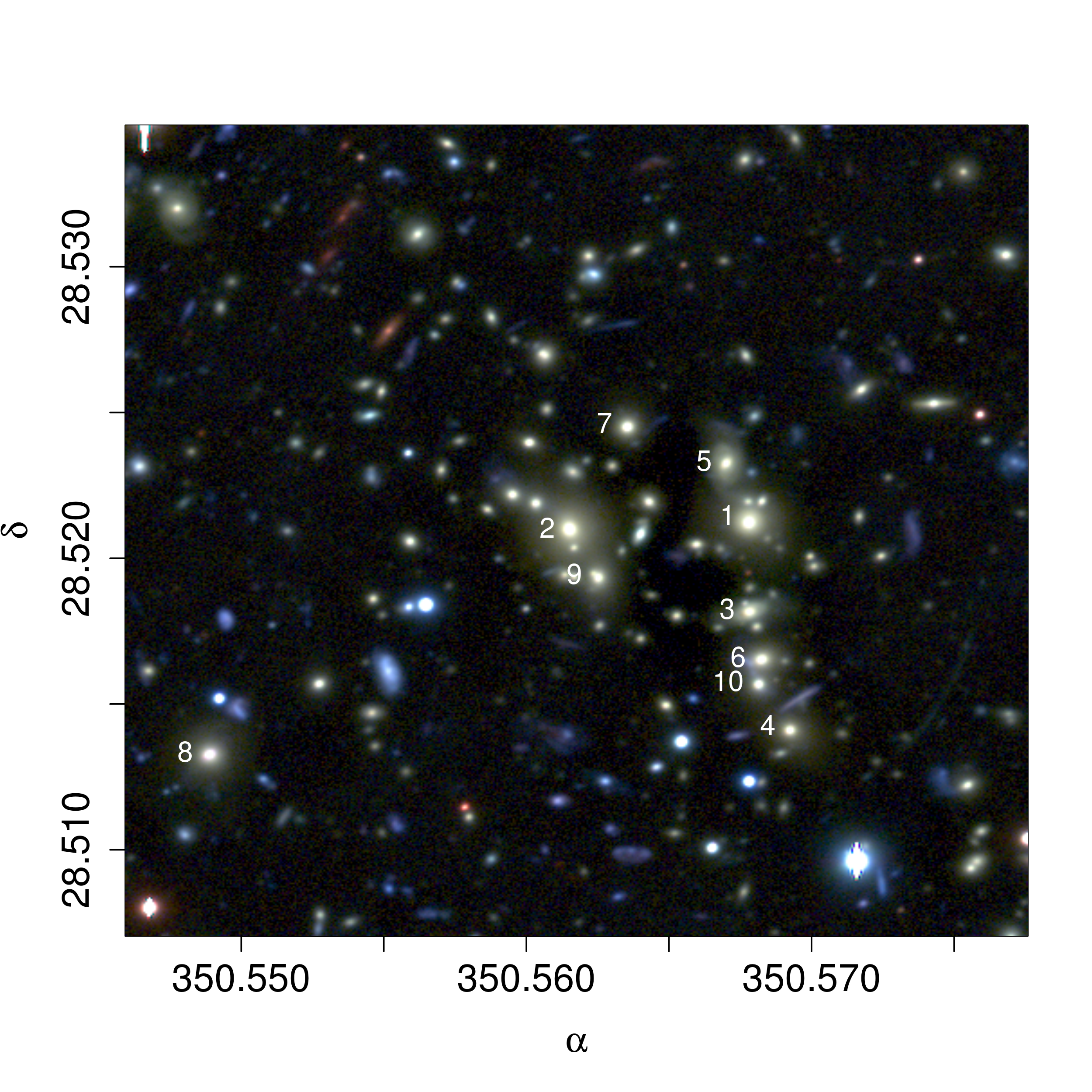}
   \caption{Zoom of the inner $100\arcsec \times 100\arcsec$ region of the PLCKG100 field. Labels mark the 10 brightest galaxies.
  Several strong lensing gravitational arcs are visible. }
   \label{fig:map}
  \end{figure}
  
The XMM-Newton snapshot observations appeared to reveal a class of massive clusters with low X-ray luminosity and disturbed morphology, which may indicate a population of massive clusters that are under-represented in X-ray surveys. Since the cluster masses in SZ surveys are estimated from the integrated Compton $Y$ parameter via scaling relations calibrated with optical or X-ray  observations, this may have an effect on the estimate of the mass function,
and thus our understanding of  cluster growth and evolution. It is therefore of
fundamental importance to study these systems  in more detail.  
    
Lensing is a powerful method to recover
the surface density field of clusters, thus allowing to trace the spatial
distribution of dark matter. Simulations have shown that  provided wide-field observations of excellent quality are available, lensing allows an accurate measurement of the mass profile of clusters \citep{2010A&A...514A..93M}, and hence the radial mass/luminosity (M/L) profile   \citep{Medez10}, and its dependence on early- / late-type 
galaxies in the cluster. Combined with X-ray observations, lensing can help to
characterize the dynamical state of the clusters, and probe deviations from
hydrostatic equilibrium which may bias the X-ray masses 
\citep{2008MNRAS.384.1567M,2010ApJ...711.1033Z}. 

In this paper we describe the first  weak lensing analysis of the cluster PLCK G100.2-30.4 (hereafter PLCKG100),  based on deep SUBARU SuprimeCam images. PLCKG100 was detected via the SZ effect in the first 10 months of the Planck survey and confirmed by XMM-Newton X--ray  observations \citep{planck2011-5.1b}. The cluster redshift  was estimated  by optical $griz$ observations performed with the 0.80-m IAC80 telescope, and through X-ray spectroscopic observations of the Fe K line, giving $z_{\rm phot} = 0.38\pm0.04$ and $z_{\rm Fe} = 0.31\pm0.04$ respectively \citep{planck2011-5.1b}. The cluster mass estimate, derived from XMM-Newton X-ray observations via iteration about the $M_{500}$ - $Y_X$ relation of \citet{2010A&A...517A..92A}, is  $M_{\rm 500} = (5.60\pm0.22) \times 10^{14}$ $M_\odot$, and the  [0.1-2.4] keV X-ray luminosity was $L_{\rm 500} = (3.36 \pm 0.08) \times  10^{44}$ erg s$^{-1}$. 

Here we use deep (1hr) $r$--band and shallower ($\sim$ 30 min)  $giz$ images obtained with  SuprimeCam at the Subaru telescope, to derive the mass of the cluster by weak lensing.
The paper is organized as follows. A summary of the observations and data reduction strategy is presented in Sect.~\ref{sec:data}. Techniques adopted for star/galaxy separation  are described in Sect.~\ref{sec:sgclass}. A comparison with cluster properties derived by optical photometry is given in Sect.~\ref{sec:properties}. The weak lensing analysis is discussed in Sect.~\ref{sec:wl}. We  attempt to refine the hydrostatic mass estimate from XMM-Newton data in Sect.~\ref{sec:xray}. Summary and conclusions are given in  Sect.\ref{sec:summary}.

A standard concordance cosmology was adopted: $\Omega_\Lambda = 0.7$, $\Omega_M = 0.3$, $H_0 = 70$ km s$^{-1}$ Mpc$^{-1}$, giving a scale of 
4.7 kpc/arcsec at the redshift of PLCK G100.2-30.4 ($z=0.36$, as derived in  Sect.~\ref{sec:properties}).

   \begin{table*}
   \begin{center}
      \caption[]{Summary of the observations. The average seeing measured on the coadded images, and limiting magnitudes for 
      pointlike sources at SNR=5,10 are also reported.\label{tab:obssum}}
         \label{ObsLog}
         \begin{tabular}{lllllll}
            \hline\hline
            Date    & Band & Number & Total time & Average seeing (arcsec) & mag($\sigma=10$) &  mag($\sigma=5$)\\
            \hline
         23/07/2012    & $g$ & 10 & 1800s & 0.7 & 24.9 & 25.6 \\
          23/07/2012   & $r$ & 20&  3600s & 0.5 & 25.5 & 26.5 \\
           23/07/2012  & $i$ & 10 & 1800s & 0.6 & 24.7 & 25.2 \\
          23/07/2012   & $z$ & 8 & 1440s & 0.6 &  23.8 & 24.7 \\
            \hline
            \end{tabular}
         \end{center}
   \end{table*}

\section{Observations and data reduction}
\label{sec:data}

PLCKG100 (Fig.~\ref{fig:field}, Fig.~\ref{fig:map}) was observed in the second part of the night on July 23, 2012 in the $griz$ bands, with the wide--field Suprime-Cam camera 
\citep{suprimecam} mounted at the 8m Subaru telescope. The camera is composed of 10 CCDs, each with $2048\times4096$ pixels and pixel scale $0\farcs207$, allowing coverage of a field of $34\arcmin \times 27 \arcmin$. 
A sequence of 180s exposures was obtained following the dithering pattern recommended for SuprimeCam. The deepest (1hr) image
was taken in the $r$-band, to be used for the lensing analysis, with optimal seeing conditions ($\sim 0\farcs5$). Images in $g$ and 
$i$ were obtained with a  total exposure time of 1800s as planned, while due to technical reasons only 8 out of the 10 planned 
exposures could be done in $z$, for a total exposure of 1440s. Details of the observations are reported in Tab.\ref{tab:obssum}.

The prereduction (overscan correction, bias, flat-fielding and masking of bad columns and autoguider) was based on the 
SDFRED2 \citep{2002AJ....123...66Y, 2004ApJ...611..660O} pipeline developed for Suprime--Cam.

The tool {\sc AstromC} \citep{Romano10} was then used to compute the astrometric solution describing the deformations in the field of view of 
individual exposures. For each filter, the astrometric solution  was computed taking the NOMAD catalog as reference,
and at the same time minimizing the differences in the position of the same sources measured in different exposures: this allowed us to obtain an internal accuracy of $\sim 0.01$ arcsec, better than the accuracy of the NOMAD catalog (rms $\sim$ 0.2 arcsec).
Images were then resampled and finally coadded using the  {\sc Swarp} software\footnote{http://www.astromatic.net/software/swarp} \citep{SWarp}.

Photometric calibration was done by observing a field selected from the Stripe 82 area, and using  the SDSS Stripe 82 Standard Star Catalog \citep{2007AJ....134..973I} to  derive photometric zero points and SuprimeCam vs. SDSS color terms. 
Limiting magnitudes for point--like sources at signal to noise levels of SNR=5,10 were computed as in \cite{virmosu}: simulated images with the same depth and background rms as each science 
image were produced using  {\sc SkyMaker}\footnote{http://www.astromatic.net/software/skymaker} software, and stars added  at random positions; errors on magnitudes were computed from the difference between input and measured magnitudes, and used to derive the median magnitude for each value of SNR. 

In addition to the stacked image obtained as described above, we also produced a set of stacks where the PSF was homogenized to the same gaussian shape for all  filters. This was done as follows:
\begin {enumerate}
\item For each exposure, all CCD images were first regridded and combined in one image using {\sc SWarp}, adopting the same center, pixel scale and size 
as the coadded $r$--band stack derived above.
\item For each of them, stars were identified  and the PSF fitted as a function of pixel coordinates using 
the  {\sc PSFex} software\footnote{http://www.astromatic.net/software/psfex} \citep{PSFex}, as described later in Sect.~\ref{sec:sgclass}.
\item The {\sc PSFMATCH} task in {\sc IRAF} was used to compute the kernel that transforms that PSF into a Gaussian with the target FWHM, chosen to be 0.9 arcsec, the worst seeing value measured in all exposures and bands: since the PSF changes across the field, we divided each 
image in sections of size $100\times100$ pixels, and computed the kernel in  each of them. 
\item For each filter, all convolved exposures were finally summed together.
\end{enumerate}

These PSF-homogenized stacks were used for the photometric measurements only, not for shape measurements, for which we used the unconvolved $r$--band image.  A multi-band 
photometric catalog was extracted by running {\sc SExtractor}\footnote{http://www.astromatic.net/software/sextractor} \citep{SExtractor} in dual mode, taking the $r$--band as the detection image, and measuring the photometry on the convolved stacks. 

The Galactic dust extinction in the field covered by Suprime-Cam was computed using the Schlegel maps \citep{schlegel}, obtaining an average value and rms:  
$E_{\rm B-V} = 0.12 \pm 0.01$. For each galaxy, observed magnitudes were corrected in each filter by adopting the dust extinction computed at its position.

\section{Star-galaxy classification}
\label{sec:sgclass}

The selection of non-saturated stars and galaxies was undertaken in the magnitude (MAG\_AUTO) vs. size ($\delta$ = MU\_MAX-MAG\_AUTO) space \citep{Huang11}. MU\_MAX is 
the peak surface brightness above background, which is constant for saturated stars. As in \citet{Huang11}, we rejected those sources with $\delta$ lower  than for stars as spurious detections.

\begin{figure}
 \centering
 \includegraphics[width=9.5 cm]{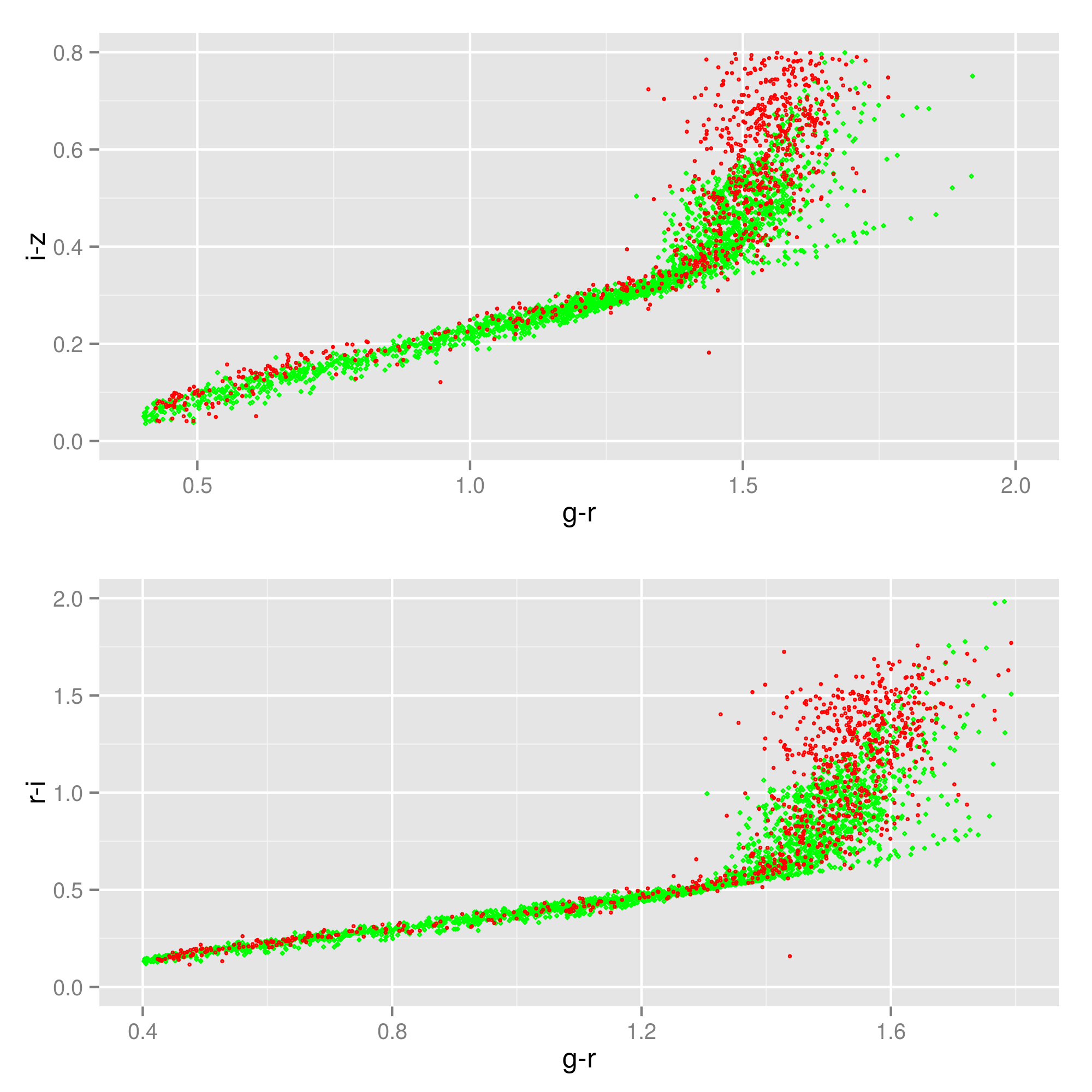}
 \caption{Stellar colors ({\em top:} $i-z$ vs. $g-r$;  {\em bottom:} $r-i$ vs. $g-r$): measured data (red dots) are compared to the values obtained by TRILEGAL (green dots).}
 \label{fig:stcols}
\end{figure}

As a test of both the photometric accuracies and of the star-galaxy classification, we compared  the $g-r$ and $i-z$ observed colors of stars  with those derived  at the Galactic plane position 
of PLCKG100 by the {\sc TRILEGAL} code \citep{2005A&A...436..895G, 2012rgps.book..165G}.  {\sc TRILEGAL}  allows the simulation of broad-band photometry of stars in any Galaxy field.  The effect of the Galaxy dust extinction on stellar colors was also included in the models, starting from the average and rms $E(B-V)$ values derived from the Schlegel maps, and assigning
to each simulated star a random value within this distribution.
The transmission curves 
obtained by the  combination of SUBARU/SuprimeCam CCD, telescope, atmosphere and filter contributions were adopted. Model and observed stellar colors are displayed in Fig.\ref{fig:stcols}.
The  zero point offsets required to  optimize  the overlap of observed and model colors are:  +0.038 mag ($g$), -0.030 mag ($r$), -0.023 mag ($i$), -0.016 mag ($z$).

Stars with SNR$>$ 100 were extracted and used 
in the {\sc PSFex} software, designed to model the PSF variations in astronomical images. The code allows to model by a polynomial the pixel-by-pixel PSF 
intensity in a stamp of a given size as a function of the position in the field of view. Starting from this model, we can build a PSF stamp at each galaxy position, 
and derive the terms needed for the PSF correction of ellipticities, as described in Sect.~\ref{sec:ksbex}.  
Different combinations of polynomial orders and stamp sizes were tried; based on the analysis of the residuals of the correction of the anisotropic PSF component (see later in Sect.~\ref{sec:ksbex}),  we selected  a polynomial of order  6 and stamps with size of 20x20 pixels ($\sim 5 \times {\rm FWHM}$).

  \section{Cluster photometric properties}
  \label{sec:properties}
  
  Figure~\ref{fig:map} displays the composite $gri$ image of the cluster field: the brightest 10 galaxies within 1 arcmin from the X--ray peak \citep{planck2011-5.1b} are marked. 
  Two galaxies of similar brightness ($r \sim $ 18.20 mag) are present in this region. We select as the center of the cluster the galaxy that lies closest to the X--ray peak, and 
  identify it as the Brightest Cluster Galaxy (BCG). Several strong--lensing  gravitational arcs are also visible, the brightest of which lies at $\sim$ 40 arcsec from the BCG. Another lies at $\sim$ 60 arcsec from the BCG.
  
  \begin{figure}
   \centering
   \includegraphics[width=9.5 cm]{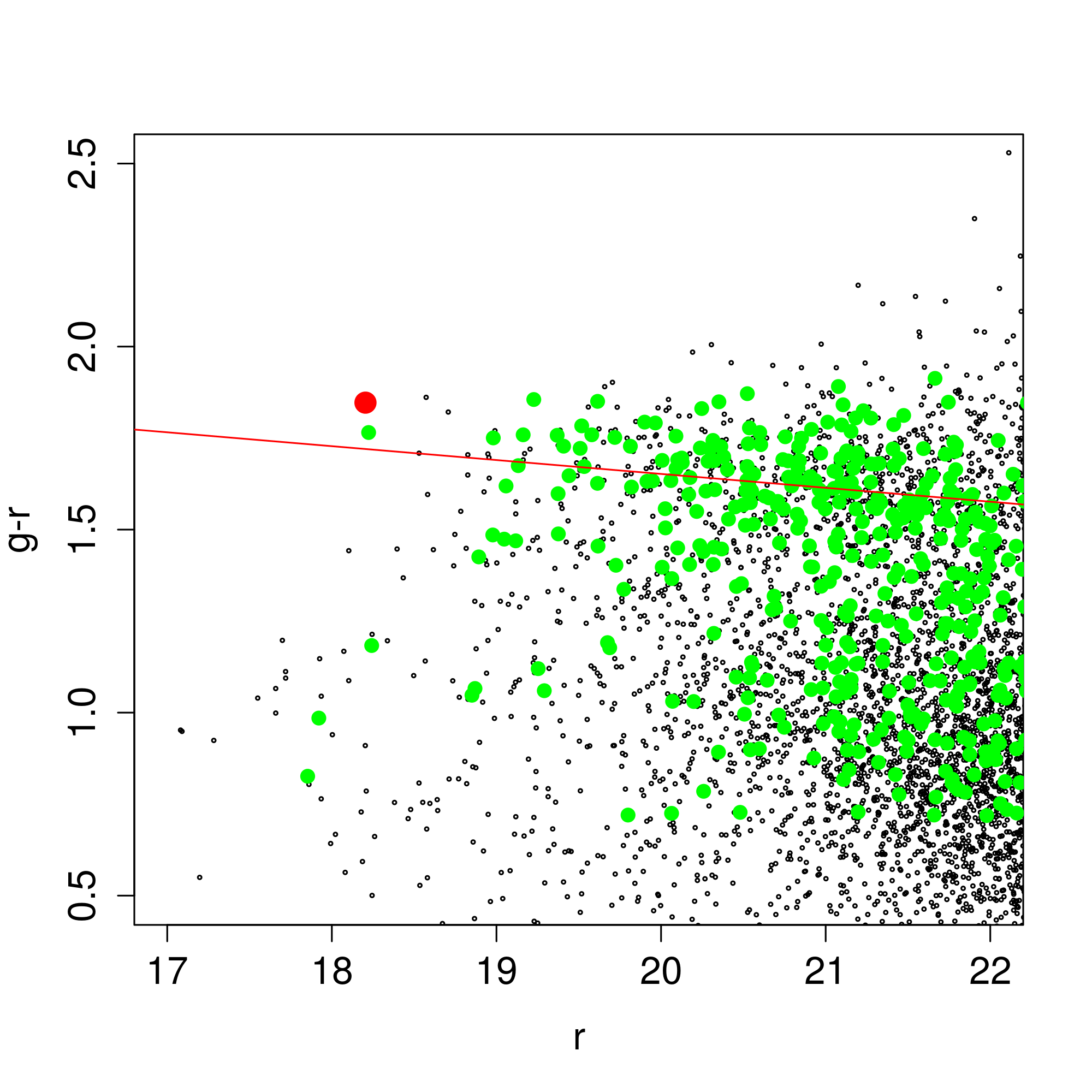}
   \caption{$r$ vs. $g-r$ color plot of galaxies in the field. Galaxies classified as likely cluster members from their colors are plotted as green dots; the red dot is the BCG galaxy. The best fitting red sequence line is overplotted.}
   \label{fig:colmag}
  \end{figure}

  \subsection{Red sequence}
  We  selected the galaxies located in the inner 5 arcmin around the BCG that were classified as cluster members  according 
  to their colors, as described later in Sect.~\ref{sec:sample}.
  The red sequence, $g-r = a + b r$, was fitted by a biweight algorithm, giving $a=2.6$, $b=-0.05$ (Fig.~\ref{fig:colmag}). Early-type galaxies were selected as those lying within $2\sigma$ from the best fitting line. 
  Photometric redshifts were computed using the  {\sc Zebra} code \citep{Feldmann06}: the cluster redshift, derived from the average photo-z of $r<22$ mag cluster members, is $z = 0.36 \pm 0.08$. For comparison, we obtain $z=0.41 \pm 0.10$ from the Sloan Digital Sky Survey (SDSS) photometric redshifts (DR9).
  
 \subsection{Richness}

  $N_{200}$, the richness of the cluster within $r_{200}$, was derived both from the selection of red-sequence cluster 
  galaxies, and via making a statistical background subtraction.  We used the $r_{200}$ derived from the
  weak lensing analysis detailed below in Sect.~\ref{sec:wl}:  $r_{200} \sim 6$ arcmin. Absolute $r-$band magnitudes and k-corrections were computed for all galaxies with the cluster redshift,
  $z=0.36$, and to a magnitude limit of  $M_{\rm abs}$ = -20 mag. The richness derived from selection of red sequence galaxies produces  $N_{200} = 150$. The richness  estimate based on statistical background subtraction was done following \citet{AA.433.415P}: the  density of background  and 
  foreground galaxies was then computed in each of 20 sectors in an annulus with thickness 0.1 deg, starting 
  at  $r_{200}$+0.1 deg. Sectors with too few galaxies  were rejected, and the average density was finally computed, giving the number of background/foreground galaxies 
  in  the $r_{200}$ area  of the cluster. In this way, we derived a richness $N_{200} \sim 135$. 
  
  The richness can be translated to a mass estimate using e.g. Eq. 16 in 
  \citet{2010MNRAS.404.1922A}:  $\log M_{200} = (0.96 \pm 0.15) (\log N_{200}-1.5) +14.36\pm0.04$.
  From these two consistent values of the richness, a mass $\log M_{200}  \sim (15.0 \pm 0.1) $ $M_\odot$ is derived.

\section {Weak lensing analysis}
\label{sec:wl}

\subsection{Shape measurement }
\label{sec:ksbex}

The derivation of the shear, and the correction of the effect of the PSF on
galaxy shapes, is based on the Kaiser Squires and
Broadhurst approach \citep[][KSB hereafter]{kaiser95}. In the last decade there has been a large effort to develop new algorithms to improve the accuracy in shape measurement \citep[see][for a comprehensive review]{2012MNRAS.423.3163K}, which is crucial in particular to derive cosmological parameters from weak gravitational lensing (cosmic shear). 
However, KSB is still widely used in the case of the analysis of massive structures such as galaxy clusters
\citep[see e.g.][]{Okabe10, 2012MNRAS.427.1298H, 2014ApJ...795..163U, 2014MNRAS.439...48A}. Our implementation follows that described in \citet{Huang11}, with some differences that are outlined below.

The KSB algorithm is based on the measurement of the following quantities  for each galaxy, derived from the source brightness moments: the source ellipticity vector\footnote{In the following, the vector
components along the $X$ and $Y$ axes are denoted as e.g. $e1$ and $e2$, respectively.}   $e_{\rm obs}$, and the {\em smear polarizability} ($P^{\rm sm}$) and {\em shear polarizability}  ($P^{\rm sh}$) tensors. 

The source ellipticity corrected for the  anisotropic component of the PSF, $e_{\rm aniso}$ , is first derived as:
\begin{equation}
 e_{\rm aniso} = e_{\rm obs} - P^{\rm sm} p,
\end{equation}
\begin{equation}
p = e^*_{\rm obs}/P^{\rm sm*},
\end{equation}
where starred terms indicate that they are derived from measurement of stars. 
This is then related to the unknown intrinsic ellipticity $e$ of the source and to the reduced shear, $g=\gamma/(1-\kappa)$,  by:
 \begin{equation}
 e_{\rm aniso} = e +P^{\gamma} g,   \label{eq:ellipticity}
 \end{equation}
where the effect of the (isotropic)  seeing is corrected by the term $P^{\gamma}$, dubbed by \citet{kaiser97} as the \textit{pre--seeing shear polarizability}:
\begin{equation}
 P^{\gamma} = P^{\rm sh} - P^{\rm sm} \frac{P^{\rm sh*}}{P^{\rm sm*}} \equiv  P^{\rm sh} - P^{\rm sm} q.
\end{equation}

The tangential and cross components of the reduced shear, $g_t$ and $g_x$ are defined as:
\begin{equation}
 \label{eq_1.0_ila}
 g_t=- g_1\cos 2\varphi -  g_2\sin 2\varphi;\\
 g_x=- g_1\sin 2\varphi +  g_2\cos 2\varphi,
\end{equation}
 $\varphi$ being the position angle of the galaxy with respect to the assumed cluster center.

The source brightness moments are weighted by a window function to suppress the outer parts of the galaxies which 
are noise dominated. As in \citet{Huang11}, the window function size was set to the value that maximizes the ellipticity signal to noise ratio ${\rm SNe}$, defined as:
\begin{equation}
 {\rm SNe}(\theta) = \frac{\int I(\theta) W(|\theta|)d^2\theta}{\sigma_{\rm sky}\sqrt{\int W^2(|\theta|)d^2\theta}}.
\end{equation}
Sources with ${\rm SNe}$ $<$ 5 were rejected from the analysis. We further excluded  those galaxies with $P_\gamma<0.1$, for which the PSF correction of the ellipticity is less reliable.

The final output of the pipeline is the vector $e_{\rm iso} =e_{\rm aniso}/P^\gamma$. Provided  that the average intrinsic ellipticity
vanishes, $\left\langle e\right\rangle  = 0$,  the average reduced shear is $\left\langle g\right\rangle  = \left\langle e_{\rm iso}\right\rangle$: it is therefore common to set $g = e_{\rm iso}$.

\begin{figure}
 \centering
 \includegraphics[width=9.5 cm]{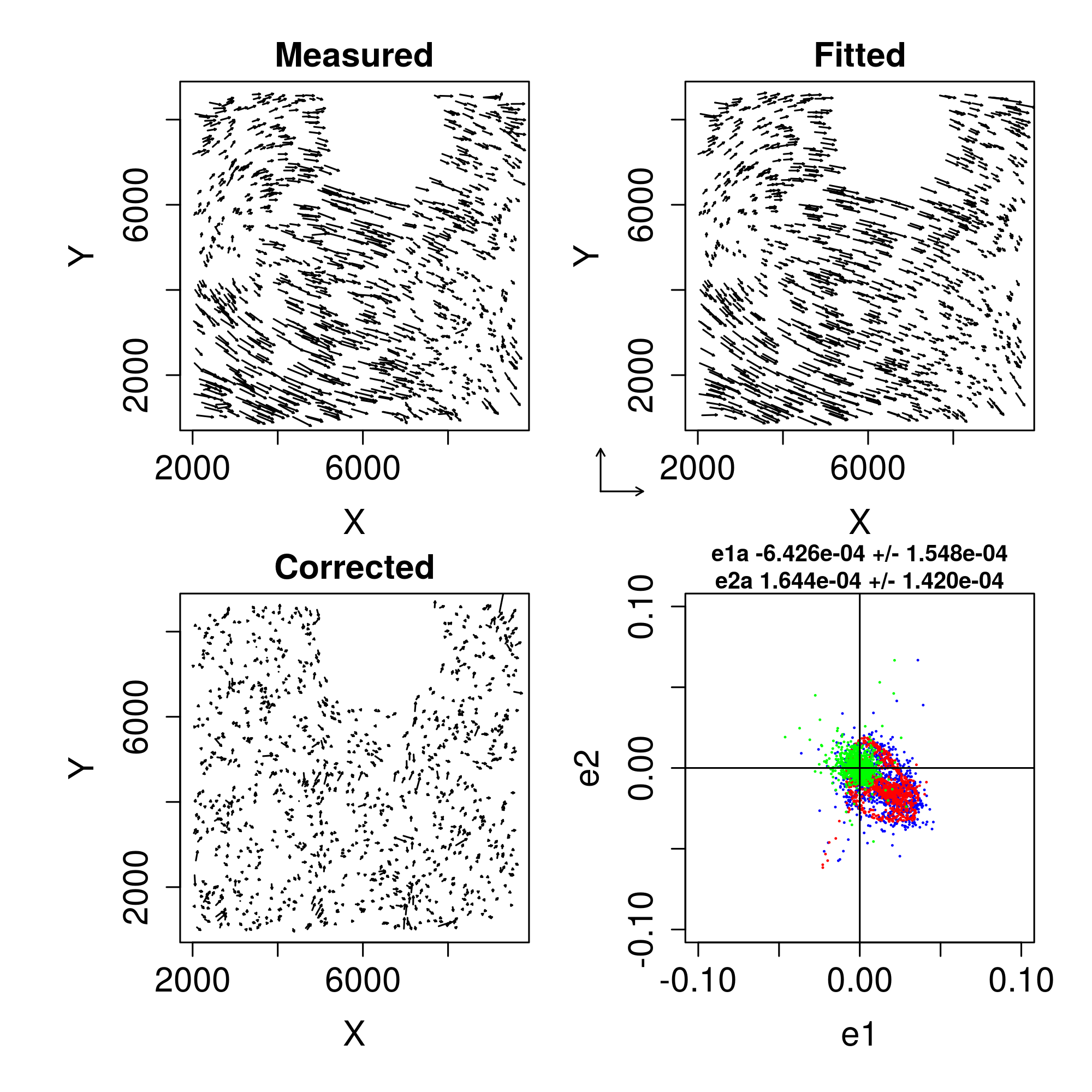}
 \caption{PSF anisotropy correction: the first three panels show the ellipticity pattern (measured, derived from the PSFex model and corrected 
($e_{\rm aniso}$), X and Y are in pixels). The scale is displayed by the arrows in the middle ($e=0.1$).
In the last panel, blue and red dots are the values of $e_{\rm aniso}$ derived on real stars and on the PSFex model respectively; green dots are the values after the correction. }
 \label{fig:psfmap}
\end{figure}

\begin{figure}
 \centering
 \includegraphics[width=9.5 cm]{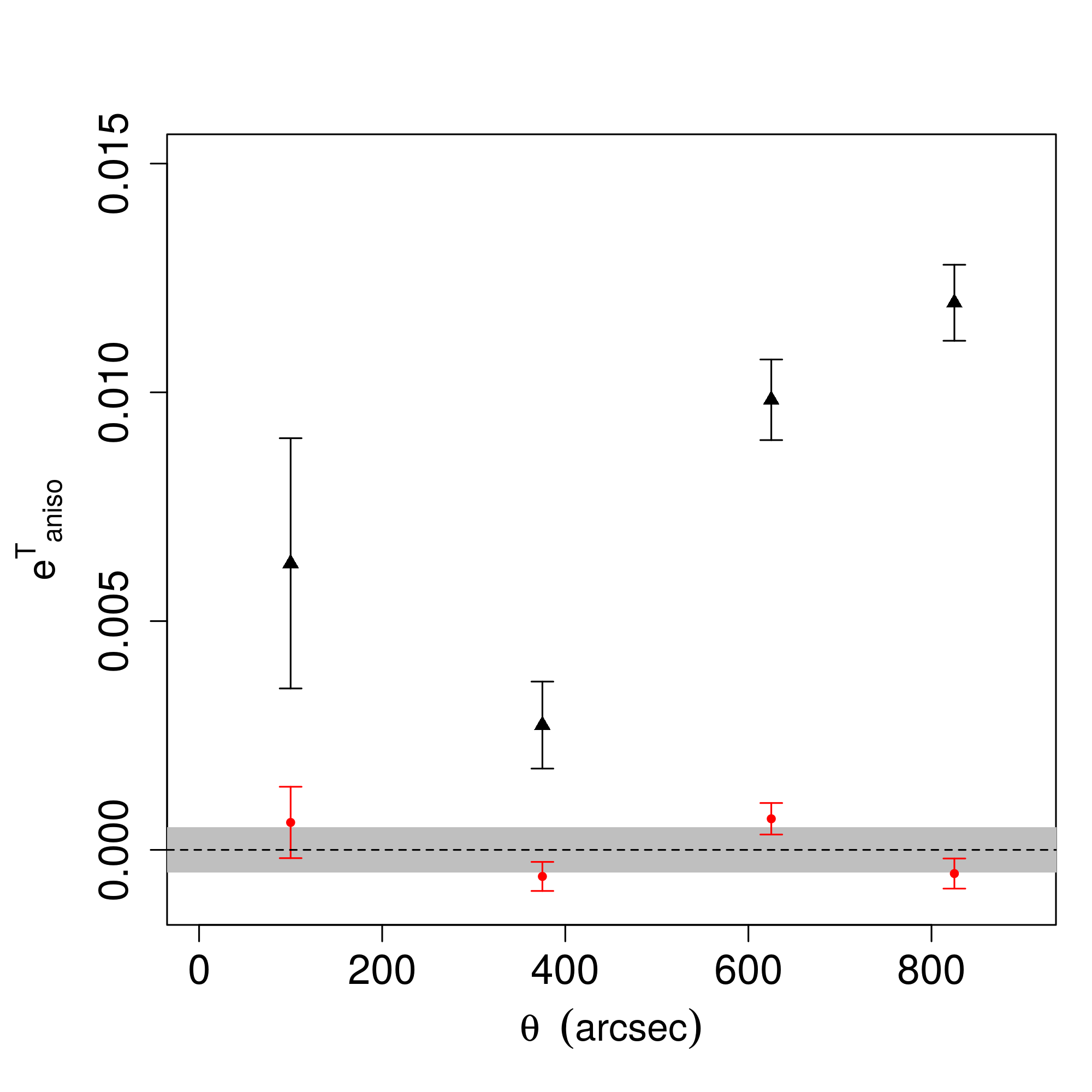}
 \caption{Average tangential PSF ellipticity at different distances from the cluster centre, before (black triangles) and after (red circles) the anisotropy correction. The shaded area is plotted at $e_{\rm aniso}^T = \pm 5 \times 10^{-4}$.}
 \label{fig:psfanisot}
\end{figure}

As first pointed out by \citet{Hoekstra98}, it is important to adopt the same size of the window function to compute the galaxy ellipticities and the stellar correction terms used
for the PSF correction. In \citet{Huang11} we computed the stellar terms in bins with $\theta$ varying between 
the minimum and maximum values allowed for the size of the galaxies (2 and 10 pixels respectively), with a step of 0.5 pixels. For each bin, they were then fitted by a polynomial, as a function of pixel 
coordinates,  so that they could be derived at the position of each galaxy. This approach presents some disadvantages: (i) it is necessary to make a polynomial fit for each PSF correction term and for each window function size; (ii) large window sizes increase the amount of noise in the measurement of moments, even if we are using bright stars, which may contribute to different biases as a function of galaxy size.   
A more recent approach consists of modelling the PSF \citep{2012MNRAS.419.2356B}, enabling us to reproduce the PSF at each galaxy's position: this is now feasible with accuracies \citep{2013ApJS..205...12K} $\sigma(e) < 10^{-3}$. To this end, here we used the \textsc{PSFex}
\footnote{http://www.astromatic.net/software/psfex} software, which allows modelling of  the PSF as a linear combination of basis vectors and fits their coefficients with a polynomial as a function of the pixel position. For each galaxy position, PSF correction terms are then derived using exactly the same window function  adopted for that galaxy. A similar approach was adopted in \citet{Gruen13}, who showed that using \textsc{PSFex} to model the PSF provides the accuracy required for weak lensing analysis of galaxy clusters.
The accuracy of the PSF fitting was checked by deriving the anisotropy corrected
ellipticities for stars, and verifying the absence of systematic effects in the residuals across the field of view (see Fig.~\ref{fig:psfmap}). Using a polynomial of order 6 as described above, we obtained $\langle e1_ {\rm aniso} \rangle  = (-6 \pm 2) \times 10^{-4}$, $\langle e2_ {\rm aniso}\rangle = (2 \pm 1) \times 10^{-4}$. Fig.~\ref{fig:psfanisot} shows the tangential component of the average stellar ellipticity before and after the anisotropic correction ($< 6 \times 10^{-4}$), at different distances from the cluster centre.  All of the steps outlined above are implemented in the \textsc{lensR} pipeline introduced in \citet{Huang11}.

For each galaxy, we adopted the same  weight definition as in  \citet{Huang11}:
\begin{equation}
w=\frac{1}{\Delta {e_0}^2  +\Delta e_{\rm iso}^2 }, \\ \label{eq:well}
\end{equation}
where $\Delta {e_0} \sim 0.3$ is the typical intrinsic rms of galaxy ellipticities. The uncertainty on the corrected ellipticities, 
$\Delta e_{\rm iso}$, was computed following a numeric approach. For each star used in the PSF fitting, we compute the difference between the 
correction terms measured on the real stars and on the PSF model derived by PSFex, and fit them as a function of the position. This allows 
estimation of an uncertainty on the stellar correction terms at each galaxy's position. $N$ values of $e_{\rm iso}$ were then computed  
(e.g. $N$=100), randomly drawing  both the observed source ellipticity and the stellar correction terms from a Gaussian distribution built from their 
errors. The error on  $e_{\rm iso}$ used to compute the weight was taken to be the rms of these values.

\subsection{The background sample}
\label{sec:sample}

The dilution bias introduced by unlensed galaxies in the sample used for the
shape measurement can adversely affect the cluster mass estimate.
To minimise the contamination of this sample by unlensed sources, we
followed the  approach described in \citet{Formicola13}. 
Using the galaxy catalog with   photometry in $griz$ bands, we identified 
background, cluster members and foreground populations in a color-color diagram
derived by choosing the appropriate combination of colors to be investigated.

Our selection in the color-color (CC) space was based on  the simultaneous
analysis of galaxy colors and of the shear signal of the galaxies selected as background sources.
We used the photometric data from the Cosmic Evolution Survey \citep[COSMOS;][]{ilbert} to train the color selection.
The accurate photometric redshifts of the sample allow a careful characterization of
the distribution of the galaxies belonging to different redshift ranges in CC space. At the same time, we tuned the color cuts to maximize the amplitude of the tangential shear component  (Eq.~\ref{eq_1.0_ila}) measured from the  background sources.

\begin{figure}[h]
  \begin{center} 
\includegraphics[viewport=55 15 500 350,scale=.5]{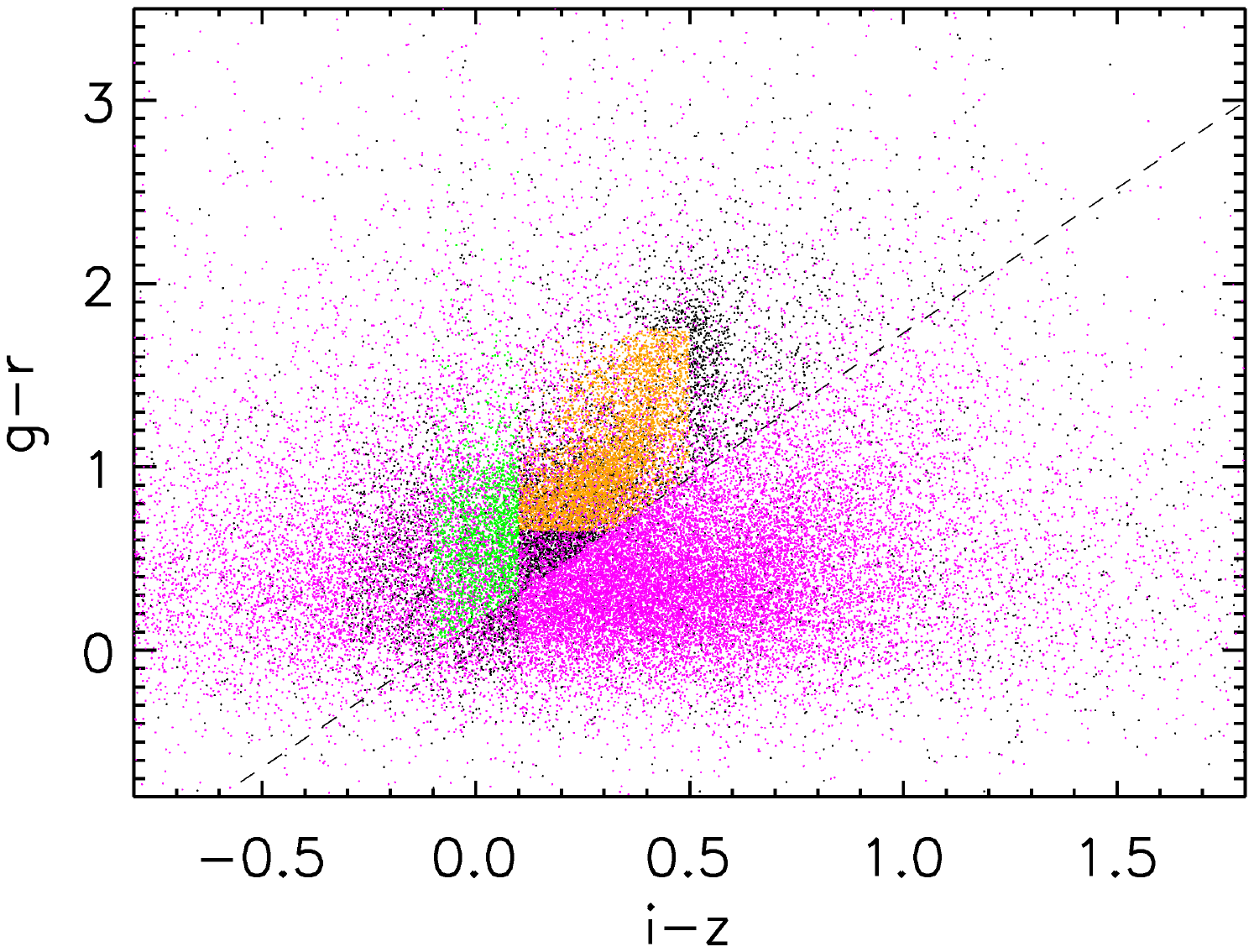}
\includegraphics[width=\columnwidth]{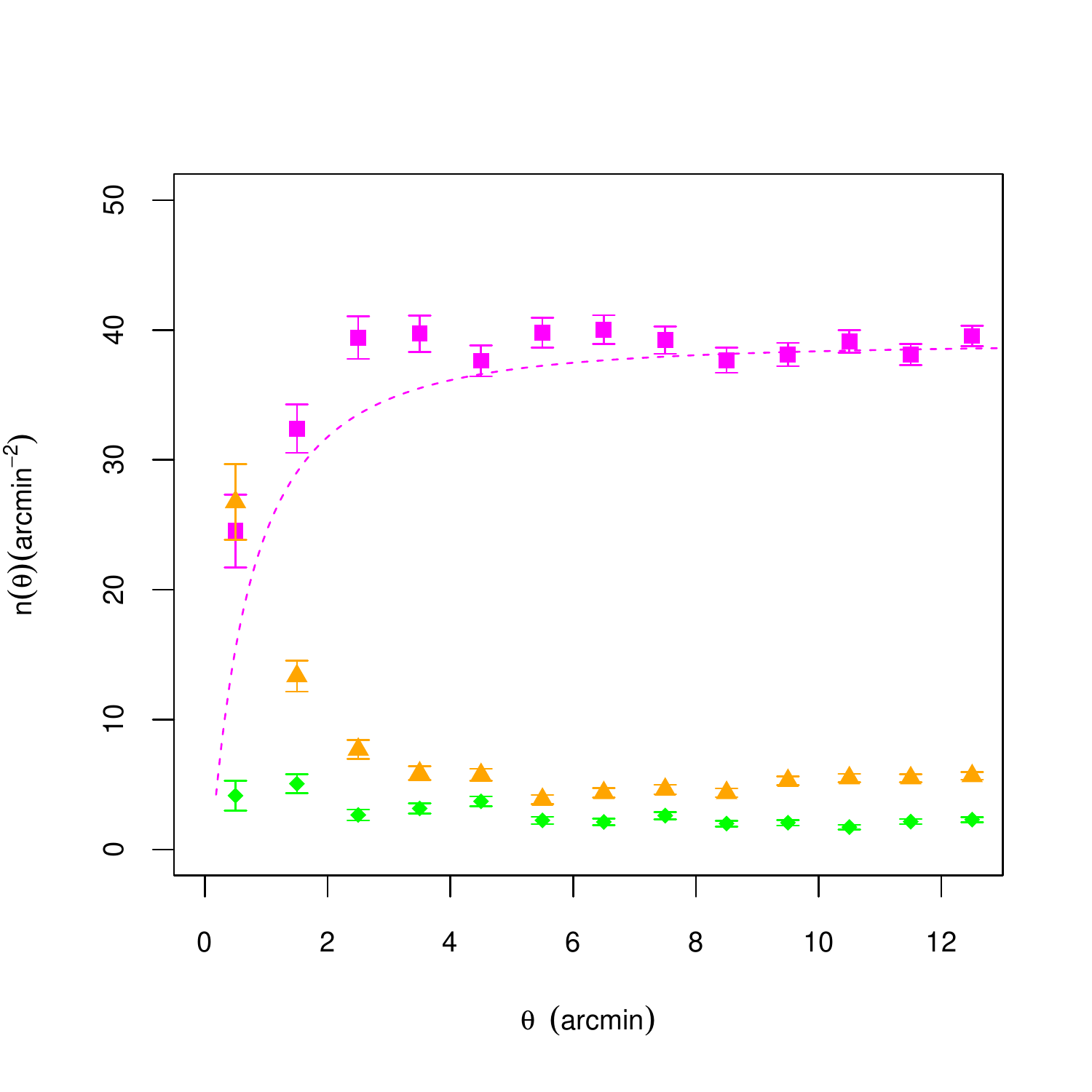}
 \caption{Selection in CC space.  Upper panel: CC diagram for PLCK G100 galaxies.
Cluster members are displayed with orange dots and  foreground and background   galaxies with green and magenta dots, respectively. Unclassified sources are plotted with black dots.
Bottom panel: radial number density profiles of galaxies. The background  density profile (magenta squares) shows a decrease in the central region; overplotted is the magnification effect expected for a NFW profile.
}
\label{col_plck}
\end{center}
\end{figure}

Starting from the $i-z$ vs $g-r$ CC plot of the COSMOS galaxies,  we defined  
criteria based on color and magnitude cuts allowing to identify each galaxy population in the PLCKG100 images. To this end, we computed the distance $d$ of each galaxy in the  $i-z$ vs $g-r$
 diagram  from the line represented in  Fig.~\ref{col_plck} (upper panel). 
 This line (derived as described in \cite{Formicola13}) separates the over-dense region
mainly populated by sources at low redshift, from the over-dense region mainly populated by sources at redshift higher than that of the cluster.  

The background population was selected according to  the following criteria:
 \begin{itemize}
 \item  galaxies with: $24.5 < r < 26.5$ mag;
 \item  or:  $d < 0$,  $21.8 < r < 24.5$ mag, 
 $i-z > 0.1$;
\item    or: $d \ge 0$, $21.5 < r < 24.5$ mag,
 $i-z <- 0.3$.
 \end{itemize}

The foreground and the cluster member population were identified according to these criteria:
\begin{itemize}
\item foreground galaxies:  $d \ge 0$, $21.0 < r < 24.5$ mag, 
   $-0.1<i-z <0.1$;
\item cluster member galaxies: $d \ge 0$,  $r < 23.5$ mag,   $0.65<g-r <2.0$, $0.1<i-z<0.5$.
\end{itemize}

The total sample of background galaxies  is represented with
magenta dots in 
Fig.~\ref{col_plck} (upper panel), 
while foreground and cluster galaxies are marked with green and orange dots,
respectively.
Radial number density profiles  for each identified population are shown in Fig.~\ref{col_plck}
(bottom panel).
The  radial number density profile  for cluster member
sources (orange triangles) indicates a clear identification of the cluster member
population as shown by the high clustering at small radii.
As expected, the foreground population (green diamonds) appears uniformly distributed.
The background sample (magenta squares)  shows  a  drop in number counts towards the cluster centre, which is expected due to the so--called magnification bias \citep{broad_2008}. Also displayed in Fig.~\ref{col_plck} is the depletion in counts expected for a NFW profile with the parameters derived by the weak lensing analysis in Sect.~\ref{sec:weaklens}, showing good consistency with the observations.

Using the photo-z derived by \citet{ilbert},  we extracted the redshift
distribution of the COSMOS galaxies with  color and magnitude cuts describing
the background population,  needed for the evaluation of the critical density in the weak lensing analysis, as described in Sect.~\ref{sec:weaklens}.

 \begin{table*}
 \begin{center}
 \caption {Best-fit NFW parameters, at different overdensities, derived taking the two brightest galaxies as centre ($n=1,2$): for each  centre, in the upper row ($a$) both $M_{\rm vir}$ and $c_{\rm vir}$ were
 taken as free parameters; in the next row ($b$) the ($M_{\rm vir}$, $c_{\rm vir}$, $z_{\rm cl}$) relation \citep{2008MNRAS.390L..64D} was used; in the last three rows  we use the $c-M-z$ 2D relations in \citet{2014ApJ...797...34M} ({\em c}: all, {\em d}: relaxed,  {\em e}: super-relaxed).
 Masses are in units of $10^{14} M_\odot$, radii in kpc.
 \label{tab:nfw}}
 \renewcommand{\arraystretch}{1.5}

 \begin{tabular}{lllllllll}
   \hline\hline
 n & $M_{\rm vir}$ & $c_{\rm vir}$ & $r_{\rm vir}$ & $M_{\rm 2500}$ & $r_{\rm2500}$ & $M_{\rm 500}$ & $r_{\rm 500}$  & $M_{\rm 200}$  \\ 
   \hline
   
   1s & $11.72_{-1.55}^{+2.35}$ & $5.98_{-1.45}^{+1.27}$ & $2120.70_{-97.70}^{+133.39}$ & $3.31_{-0.43}^{+0.48}$ & $518.27_{-23.35}^{+23.74}$ & $7.39_{-0.78}^{+1.50}$ & $1158.17_{-42.43}^{+73.45}$ & $10.25_{-1.26}^{+1.95}$ \\ 
     1b & $15.31_{-2.00}^{+2.43}$ & $3.77_{-0.04}^{+0.04}$ & $2318.68_{-105.61}^{+116.58}$ & $2.85_{-0.34}^{+0.41}$ & $493.32_{-20.61}^{+22.64}$ & $8.43_{-1.07}^{+1.30}$ & $1210.04_{-53.61}^{+59.09}$ & $12.91_{-1.67}^{+2.03}$ \\ 
     1c & $15.46_{-1.74}^{+2.85}$ & $3.65_{-0.02}^{+0.03}$ & $2326.07_{-90.41}^{+134.79}$ & $2.78_{-0.33}^{+0.54}$ & $489.09_{-20.03}^{+29.93}$ & $8.41_{-0.96}^{+1.58}$ & $1209.20_{-47.83}^{+71.36}$ & $13.00_{-1.46}^{+2.41}$ \\ 
     1d & $15.04_{-2.59}^{+2.47}$ & $3.77_{-0.02}^{+0.02}$ & $2304.89_{-140.47}^{+119.99}$ & $2.80_{-0.47}^{+0.44}$ & $490.08_{-28.90}^{+24.63}$ & $8.27_{-1.41}^{+1.34}$ & $1202.59_{-72.51}^{+61.89}$ & $12.68_{-2.17}^{+2.08}$ \\ 
     1e & $14.47_{-2.12}^{+2.66}$ & $4.28_{-0.04}^{+0.04}$ & $2275.06_{-116.91}^{+131.45}$ & $3.07_{-0.43}^{+0.53}$ & $505.21_{-24.70}^{+27.66}$ & $8.30_{-1.20}^{+1.50}$ & $1204.05_{-60.88}^{+68.39}$ & $12.34_{-1.80}^{+2.25}$ \\ 
     2a & $11.21_{-1.94}^{+1.98}$ & $6.15_{-1.25}^{+1.91}$ & $2089.44_{-128.58}^{+116.37}$ & $3.24_{-0.40}^{+0.32}$ & $514.43_{-22.41}^{+16.38}$ & $7.12_{-1.05}^{+1.00}$ & $1143.87_{-59.25}^{+51.11}$ & $9.83_{-1.63}^{+1.62}$ \\ 
     2b & $14.18_{-1.35}^{+2.99}$ & $3.80_{-0.06}^{+0.03}$ & $2259.80_{-73.98}^{+149.04}$ & $2.66_{-0.23}^{+0.51}$ & $481.84_{-14.46}^{+28.99}$ & $7.82_{-0.72}^{+1.60}$ & $1180.16_{-37.58}^{+75.59}$ & $11.96_{-1.13}^{+2.51}$ \\ 
     2c & $15.23_{-2.49}^{+2.61}$ & $3.65_{-0.03}^{+0.03}$ & $2314.16_{-133.77}^{+125.39}$ & $2.74_{-0.47}^{+0.50}$ & $486.45_{-29.61}^{+27.83}$ & $8.28_{-1.38}^{+1.45}$ & $1202.89_{-70.75}^{+66.38}$ & $12.80_{-2.10}^{+2.21}$ \\ 
     2d & $15.00_{-2.00}^{+2.38}$ & $3.77_{-0.02}^{+0.02}$ & $2302.54_{-106.99}^{+115.82}$ & $2.79_{-0.36}^{+0.43}$ & $489.60_{-22.01}^{+23.78}$ & $8.25_{-1.09}^{+1.29}$ & $1201.38_{-55.22}^{+59.74}$ & $12.64_{-1.68}^{+2.00}$ \\ 
     2e & $14.11_{-2.07}^{+1.45}$ & $4.29_{-0.02}^{+0.04}$ & $2256.27_{-115.92}^{+74.78}$ & $2.99_{-0.42}^{+0.29}$ & $501.23_{-24.48}^{+15.77}$ & $8.10_{-1.17}^{+0.82}$ & $1194.27_{-60.37}^{+38.93}$ & $12.04_{-1.75}^{+1.23}$ \\ 
   
     \hline
 
 \end{tabular}
 \end{center}
 
 \end{table*}

  \begin{figure}
   \centering
   \includegraphics[width=9.5 cm]{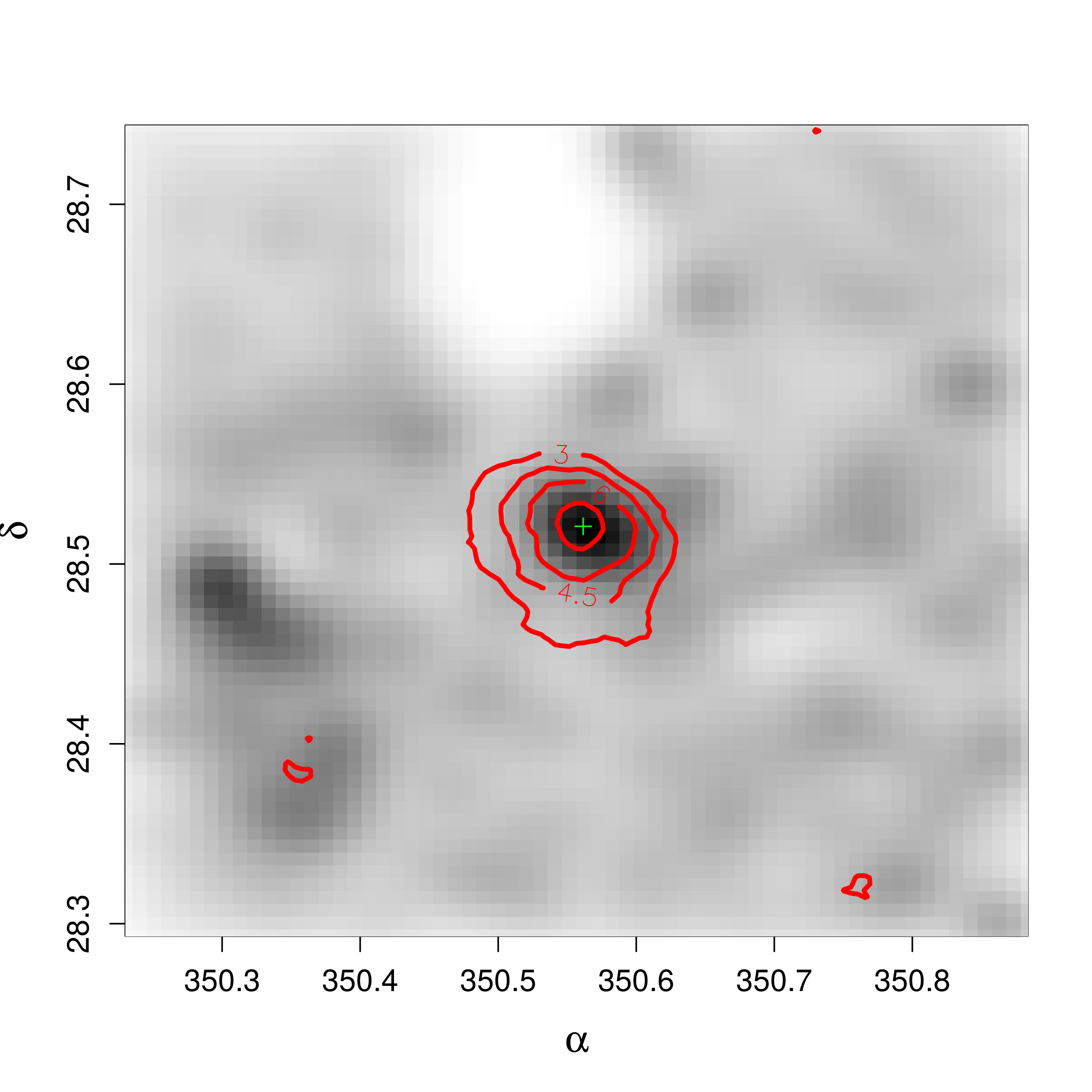}
   \caption{$r-$band luminosity-weighted  map in the PLCKG100 field. The overplotted contour lines are  the weak lensing \textit{S-map} showing the mass distribution derived by weak lensing, at levels $\sigma=$ 3, 4.5, 6, 8.5. The cross marks the position of the BCG assumed as the centre of the halo.}
   \label{fig:lumden}
  \end{figure}
  
   \begin{figure}
    \centering
   \includegraphics[width=8 cm]{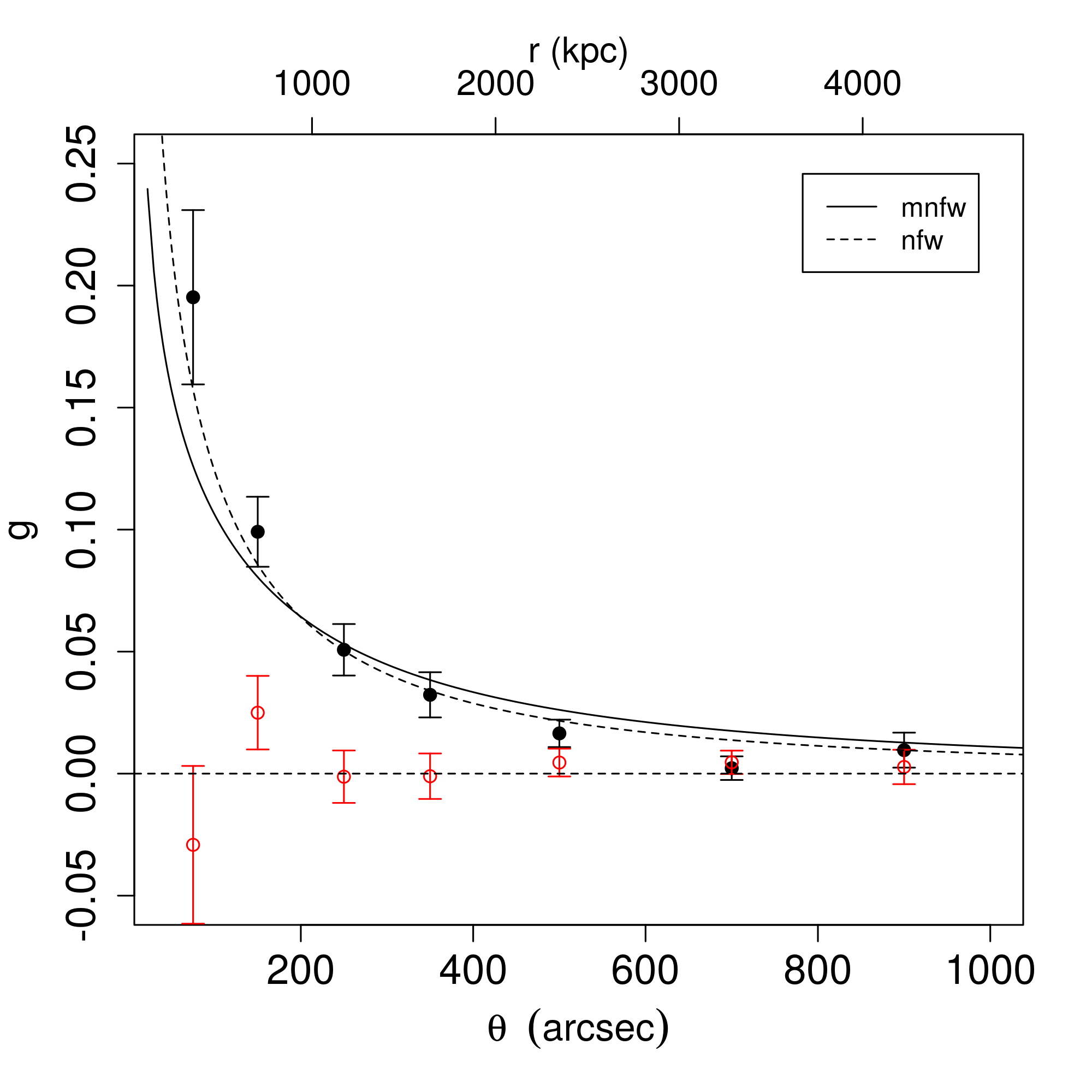}
    \caption{Shear profile detected for PLCKG100 (tangential and  cross components, where the latter is expected to be null), and 
    for comparison the values derived adopting the best-fit models. We indicate with mnfw  the model obtained using the \citet{2008MNRAS.390L..64D} relation and with nfw the model obtained keeping both $M_{vir}$ and $c_{vir}$ as free parameters.}
    \label{fig:shprof}
   \end{figure}

    \begin{figure}
      \centering
      \includegraphics[width=8.5 cm]{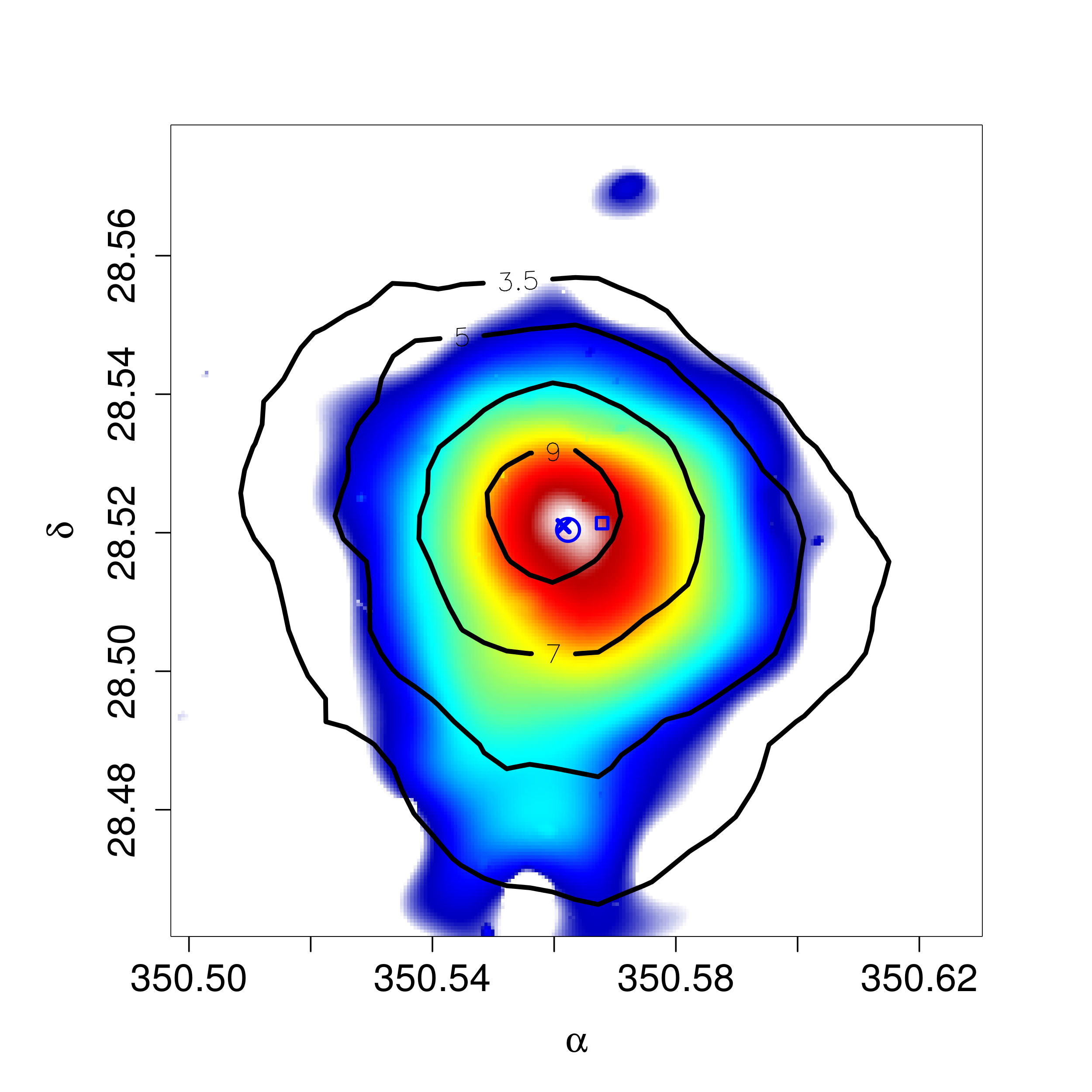}
      \caption{Wavelet reconstructed X-ray image in the $[0.5-2.5]$ keV energy band,  obtained from the combination of the three XMM EPIC camera images. The contours show  the weak lensing \textit{S-map} (see Fig.\ref{fig:lumden}), at levels  $\sigma=3.5,5,7,9$ levels. The circle is the cluster centroid derived by the XMM analysis. The  cross and square mark the position of the two brightest cluster galaxies ($1$ and $2$ respectively in the text).}
      \label{fig:xmm}
     \end{figure}

\subsection{Simulations}
\label{sec:simulations}

The ellipticities derived by most of the
available shape measurement algorithms are affected \citep[see][and references therein]{step1} by a multiplicative ($m$) and an additive ($c$) bias defined as: $e_{\rm obs} -e_{\rm true} = m\  e_{\rm true} + c$.
The value of $m$  depends on the PSF type, as well as on the brightness and size
of the galaxies, and in the case of KSB it is typically  around 10\%-20\% \citep{step1}.

To investigate this, we produced simulations using the GalSim\footnote{https://github.com/GalSim-developers/} software \citep{2014arXiv1407.7676R}, that was specifically designed to simulate sheared galaxies within the 
{\em GREAT} effort \citep[see][and references therein]{great10}. GalSim offers the possibility to use both analytic 
models for the galaxies, and/or a sample of 26\,0000 real galaxies from Hubble Space Telescope COSMOS data. In addition, it can take as input  the model produced by PSFex, and 
hence can produce an image with the same PSF shape and spatial variations as the real image. We built our simulations so that ellipticities were measured on individual stamps of sheared galaxies, in order to avoid the effect of blending from neighbour sources.

To evaluate the bias, we first produced images with constant shear values, $g =
0.01, 0.05, 0.1, 0.15, 0.2$, for each ellipticity component. We randomly drew
10\,000 galaxies from the full sample, and repeated the 
simulation 10 times. For each shear value, we then verified that the expected
input value could be recovered and evaluated the bias as a function of signal to noise ratio, as well as at different  positions  in the image to check for systematics related to a poor PSF correction.
From these tests, we derived a multiplicative bias  $m=10\%$, and no significant additive bias; we were not able to detect a significant dependence on the signal to noise ratio.

Secondly, we used GalSim to simulate the shear signal produced by  a cluster having a NFW 
profile with redshift, virial mass ($M \sim 10^{15}$ h$^{-1}$ $M_\odot$) and concentration parameter 
($c_{\rm 200} =4$) similar to what we find here for PLCK-G100. We made ten such simulations and fitted
the results in the same way as in Sect.~\ref{sec:weaklens}, where the only free parameter was the mass and the 
concentration parameter was set to the input value of the simulation. The mass obtained applying the multiplicative bias was
$M_{\rm 200} = (1.0\pm 0.2) \times 10^{15}$ h$^{-1} M_\odot$.

       \begin{figure*}
      \begin{tabular}{cc}
      
         \centering
      
         \includegraphics[width=8 cm, viewport=0 35 575 545]{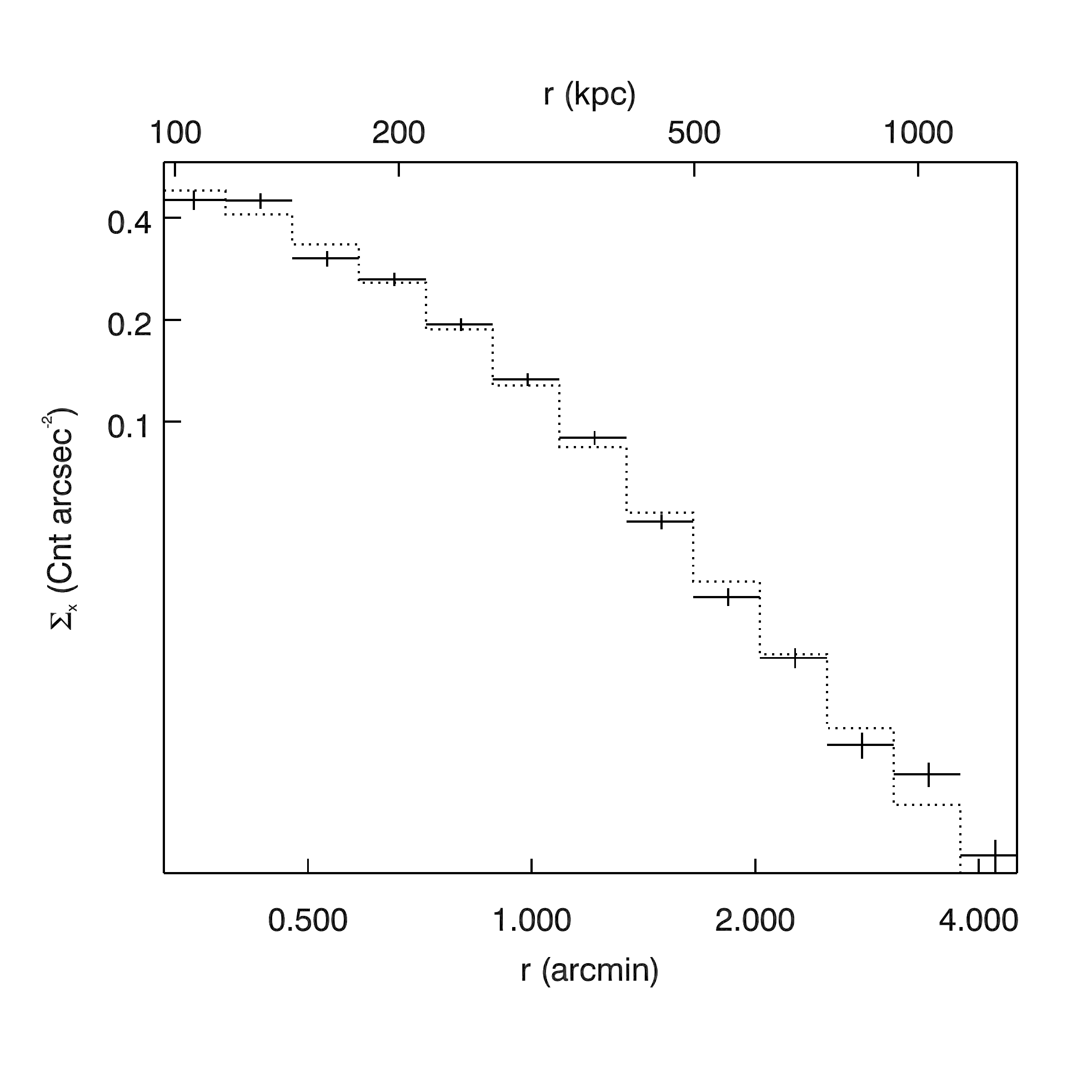} &
          \includegraphics[width=8 cm, viewport=0 35 575 545]{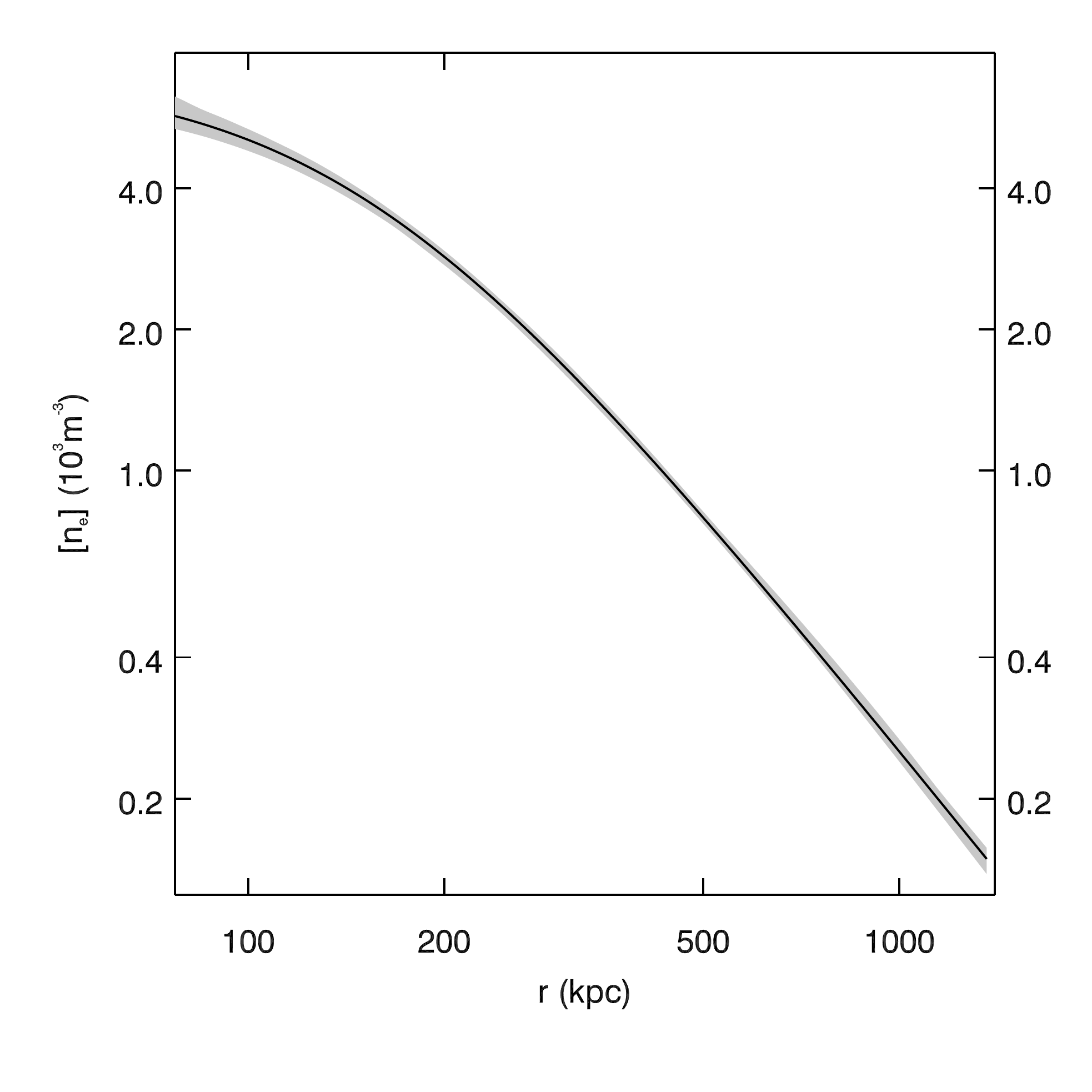} \\
            \includegraphics[width=8 cm, viewport=0 35 575 545]{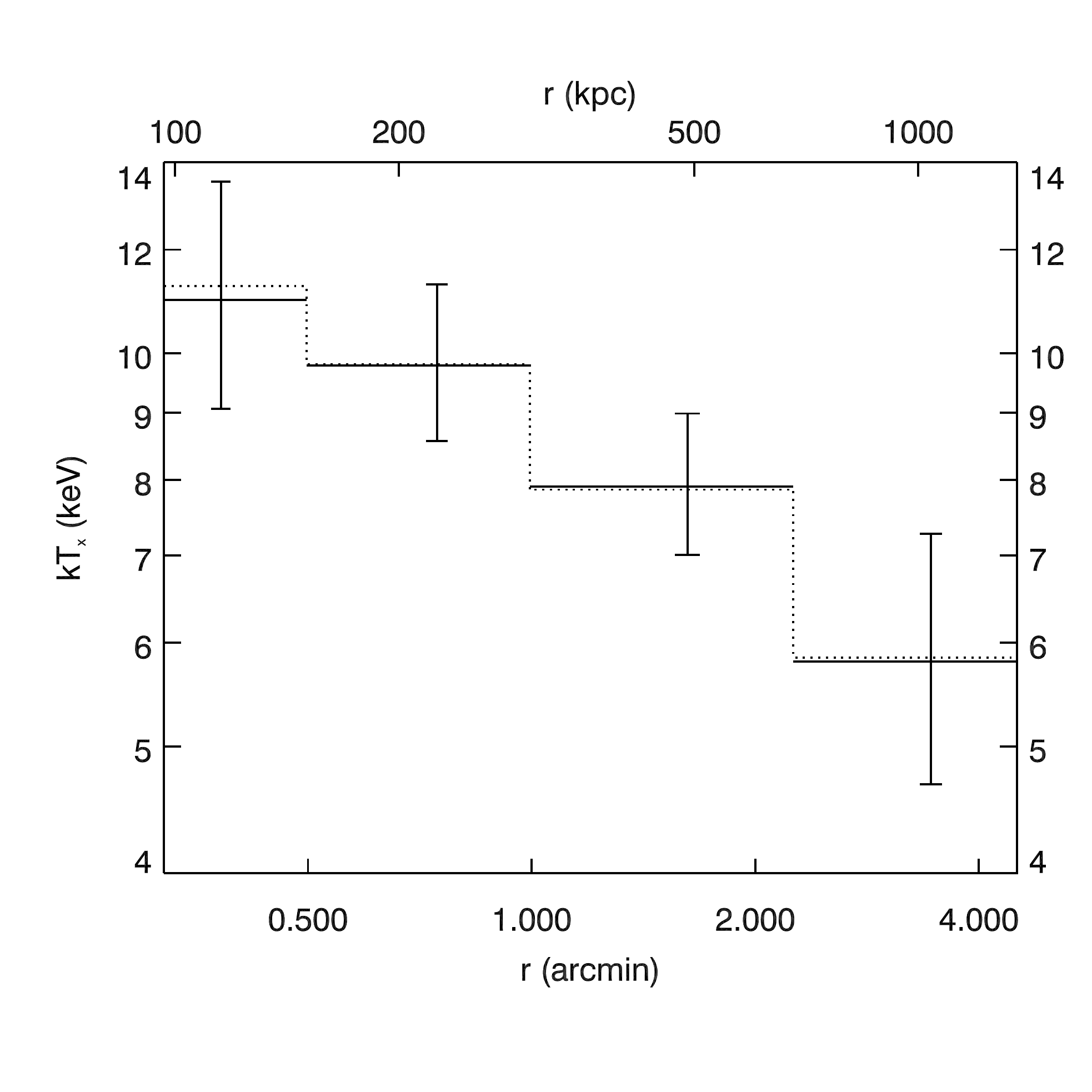} &
               \includegraphics[width=8 cm, viewport=0 35 575 545]{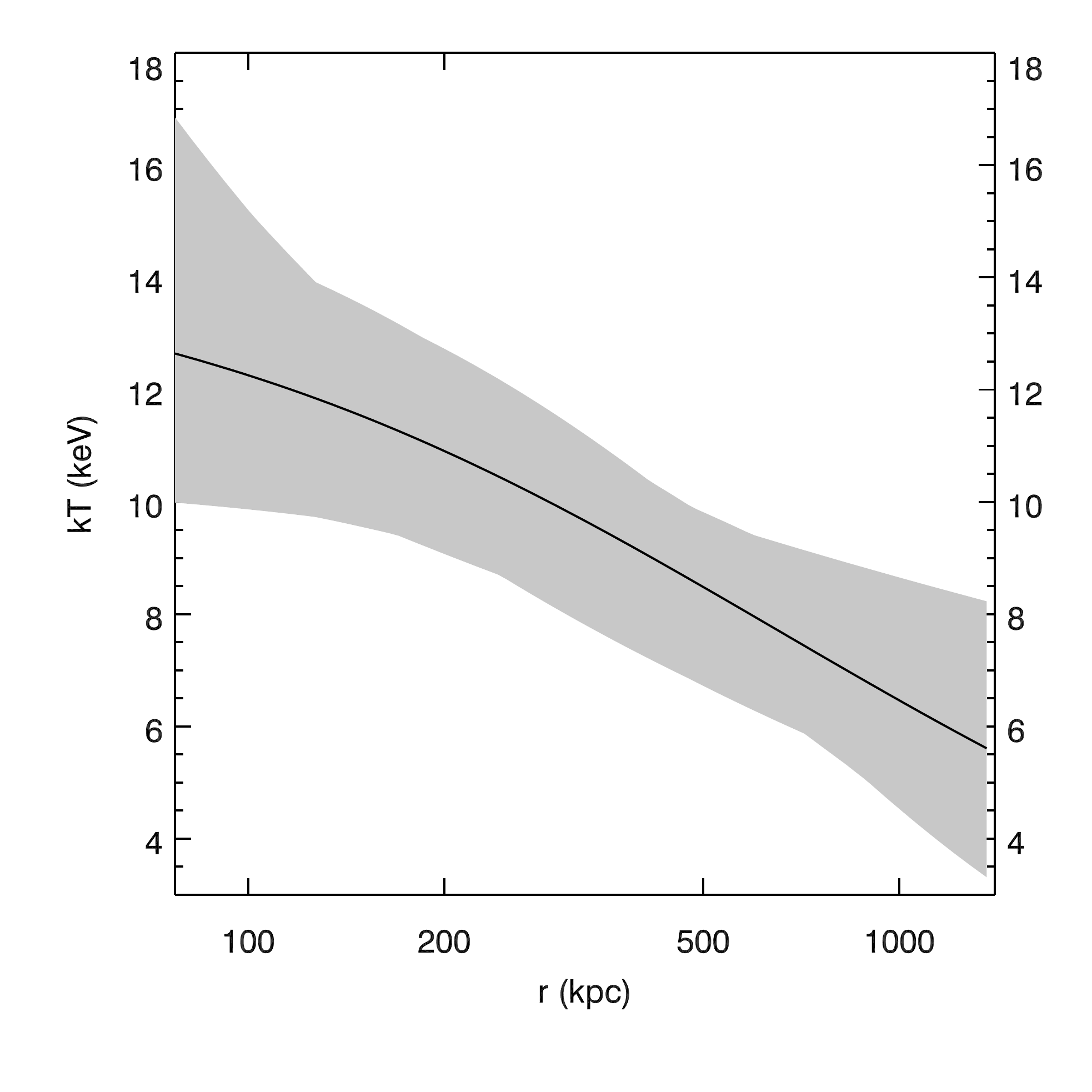}  
                      \\ 
          
        \end{tabular}
         \caption{The radial profiles of the X-ray properties are displayed as follows. 
         {\em Top, left:} surface brightness  in the [0.5-2.5] keV range. Points and the solid line represent the measured  data and the best--fit modified beta profile, respectively.
         {\em Top, right:} electron density. Solid line and grey area represent the best fit model and the relative error, respectively.    
          {\em Bottom, left:} measured temperature profile. Points and solid line represent measured data and the temperature profile, respectively.
          {\em Bottom, right:} 3D temperature profile.  Solid line and grey area represent best fit and the error, respectively.}
         \label{fig:xprops}
          \end{figure*}

\subsection{Mass estimate}
\label{sec:weaklens}

For the weak lensing analysis we used the sources in the catalog satisfying the following criteria: (i) classified as extended (Sect.~\ref{sec:sgclass});
(ii) with $P_\gamma > 0.1$, SNRe>10; e1iso, e2iso $<$ 1 ; (iii) outside regions masked due to bad columns, haloes or spikes nearby bright, saturated stars. We further considered only
those sources classified as background galaxies in Sect.~\ref{sec:sample}.
This produced a catalog of 17\,000 sources in a 0.227 sq. deg area, corresponding to a density of 21 gals arcmin$^{-2}$. 

We first extracted (Fig.~\ref{fig:lumden}) the S-map \citep[see][and references therein]{Huang11}, showing the spatial distribution of the weak lensing signal.
The peak is consistent with the position of the galaxy that we identified as the BCG of 
the cluster in Sect.~\ref{sec:properties}. We therefore took this position as the center of the cluster, and used it to compute the cross and tangential components of the 
ellipticities whose profiles are displayed in Fig.~\ref{fig:shprof}.

Computing the mass from the shear requires an estimate of 
the critical surface density:
\begin{equation}
\Sigma_{\rm crit} = c^2 (4\pi G D_l\beta)^{-1},
\end{equation}

where $\beta=D_{ls}/D_s$; $D_{ls}$, $D_s$, and $D_l$ are the angular distances
between lens and source, observer and source, and observer
and lens, respectively. Since we have no spectroscopic
training set, we are not able to judge the reliability of our
photometric redshifts for the lensed background galaxies.
We therefore adopted the  approximation of a constant $\beta$ for all galaxies.
This introduces an overestimate of the average shear  by a factor 
$1 +\left(\left\langle\beta^2\right\rangle/\left\langle\beta\right\rangle^2 -1\right) \kappa$  $\sim 1 + 0.2 \kappa$ at $z = 0.36$, as discussed in \citet{apJ...532...88h}.
The correction of this bias is included in our mass estimate.  
As described in Sect.~\ref{sec:sample}, the value of $\left\langle  \beta \right\rangle $ was computed using the photometric redshifts from the COSMOS sample, applying the color and magnitude cuts adopted for the selection of background galaxies. We find  $\left\langle \beta \right\rangle= 0.54$,  equivalent to assuming that all galaxies lie at the same redshift $z_s = 1.04$.

The shear profile was first fitted assuming a NFW halo and using  the analytical expression for the shear given by  \citet{Wright00}: a  Maximum Likelihood fitting approach was adopted \citep{Radovich08,Huang11}. 
In order to see how the mass estimate  is affected by the assumption on the cluster centre and on the model fitting parameters, we performed different tests: ({\em a}) selecting both of the two brightest galaxies as the cluster centre ($n=1,2$), and ({\em b})  trying both a 
2-parameter fit (virial mass $M_{\rm vir}$ and concentration $c_{\rm vir}$), and by assuming a relation between $M_{\rm vir}$, $c_{\rm vir}$ and the 
cluster redshift $z_{\rm cl}$. 
In the latter case, we first  used the relation in  \citet{2008MNRAS.390L..64D}:
\begin{equation}
c_{\rm vir} = A \left(\frac{M}{M_\star}\right)^B (1+z_{\rm cl})^C,
\label{eq:cvirmass} 
\end{equation}
with $M_\star= 2. \times 10^{12}/h$ $M_{\odot}$, $A = 7.85$, $B=-0.081$, $C=-0.71$. 
Then, we adopted the $c-M-z$ relations discussed in  \citet[][see their Table~2]{2014ApJ...797...34M}. These were based on numerical simulations of halos at different redshifts and halo relaxation states, undertaken for the analysis of the CLASH cluster sample. We used their 2D relations derived for NFW models for all halos, relaxed halos and super-relaxed halos. The results are given in Table~\ref{tab:nfw}. The differences in the fitted parameters are within the uncertainties, with a slightly higher mass derived for $n=1$.

A model-independent estimate of the mass was then derived using the aperture densitometry method \citep[see][and references therein]{Huang11}. 
This method provides an estimate of the projected 2D mass within a radius $\theta_m$, $M_{2D}(\le \theta_m)$. As discussed e.g., in \citet{Okabe10}, before comparing the aperture (2D)  and the NFW (3D) mass we  first need to evaluate the correction factor  $f(\theta_m) = M_{2D}(\le \theta_m)/M_{3D}(\le \theta_m) > 1$. To this end, we used again the analytical expression given by \citet{Wright00} to compute $f(\theta_m)$
for an NFW halo with the best--fit values of $c_{\rm vir}$, $M_{\rm vir}$ found above (first row in Tab.~\ref{tab:nfw}). We obtained $f=1.39$ at $r_{500}$ and $f=1.31$ at $r_{\rm vir}$. The deprojected mass values are: $M^{\rm ap}_{500} = (6\pm2) \times10^{14}$ $M_\odot$ and $M^{\rm ap}_{\rm vir} = (1.0\pm0.4)\times 10^{15}$ $M_\odot$.

The mass derived by weak lensing can be affected by the presence of large-scale structures along the 
line of sight \citep{Hoekstra03, Hoekstra11}. According to Figs. 6 and 7 in \citet{Hoekstra03}, this uncertainty
is of the order of 
$\sim 2 \times 10^{14} h^{-1}$ $M_\odot$ for a cluster with $M=10^{15} M_\odot$ at $z \sim 0.4$, 
$\theta_{\rm max} = 15$ arcmin, 
 which is comparable to the uncertainties derived in the fitting listed in Table~\ref{tab:nfw}.

   \begin{figure}
     \centering
     \includegraphics[width=8.5 cm]{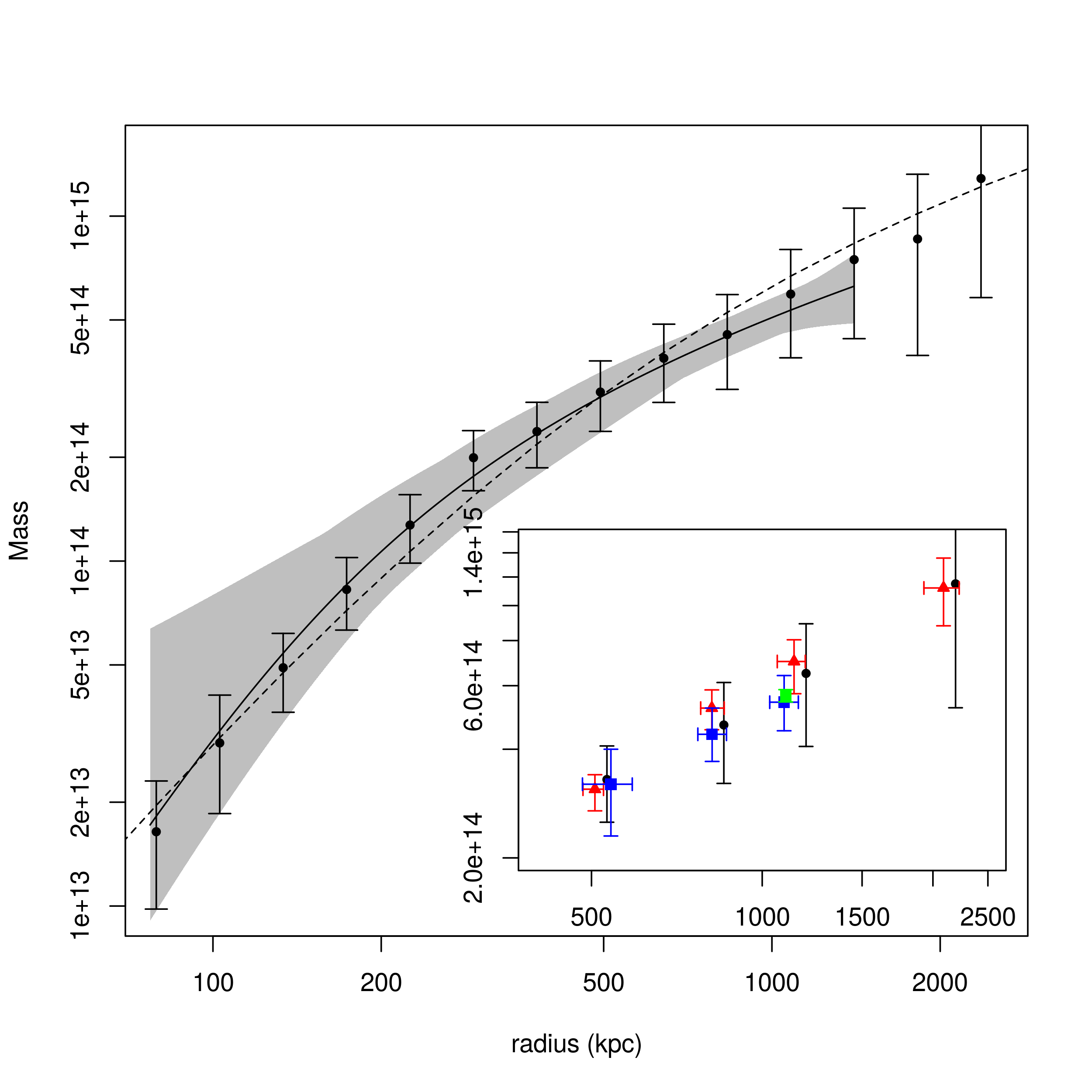}
     \caption{Radial mass profile derived from the X-ray data, where the hydrostatic mass profile and uncertainties are displayed by the solid line and the  grey area. Error bars represent the 1 $\sigma$ uncertainties described in the text. Also displayed are the mass profiles derived from the WL analysis (dashed line: NFW fit;  points with error bars: aperture densitometry after deprojection). The inset shows $M_{2500}$, $M_{1000}$, $M_{500}$ and  $M_{\rm vir}$  derived from our XMM analysis (blue squares); the Planck analysis (green square); WL NFW, first row in Tab.~\ref{tab:nfw} (red triangles); WL, aperture densitometry (black circles): a small offset in radius is applied here for display purposes. }
     \label{fig:radmass}
    \end{figure}

\section {X-ray analysis}
\label{sec:xray}

PLCK G100.230.4 has been observed for $\SI{13.7}{\kilo\second}$ with the XMM-Newton (obs-ID $0656202301$) European Photon Imaging Cameras (EPIC), during the follow-up campaign of Planck newly detected clusters. This dataset was re-processed using the XMM-Newton Science Analysis System (SAS) version $14.0.0$ in order to select the relevant calibration files. To remove any flare contamination, i.e. observation periods with an unusual count rate, we applied a 3$\sigma$ clipping and a temporal wavelet filter to the histogram of the photon arrival times, with a bin size of $\SI{50}{\second}$, in the energy band $[1-5]$ and $[10-12] \si{\kilo\electronvolt}$. After the flare removal procedure we find a useful exposure time of $\SI{13.7}{\kilo\second}$ and $\SI{5.2}{\kilo\second}$ for the two EPIC-MOS and the EPIC-PN datasets, respectively. To prepare the following imaging and spectroscopy analyses, we rebinned the flare-cleaned list of photon events in sky coordinates and energy, and associated both a 3D effective exposure and a background noise model to this data cube (see details in \citealt{mass_profiles}).

The background noise was modelled after subtraction of all detected point-like sources using two components: the instrumental noise caused by high energy particles interacting with the detectors, and several astrophysical components, namely the Cosmic X-ray background (CXB) (e.g. \citealt{giacconi}) plus a diffuse emission of Galactic origin (e.g. \citealt{snowden_gala}). For the instrumental component we used an analytical model inspired by the works of \cite{kuntz_particle} and \cite{xmm_model_lec_mol}, which holds for each EPIC camera a continuum plus several fluorescence lines (see details in \citealt{bourdin_a521}). As proposed by \cite{kuntz_gala}, the Galactic foreground was modelled using two absorbed APEC (\citealt{apec_code}) spectra, with temperatures of $\SI{0.14}{\kilo\electronvolt}$ and $\SI{0.248}{\kilo\electronvolt}$, respectively. The unresolved CXB spectrum follows a power law with $\Gamma = 1.42$ (\citealt{lumb_power}). The particle background was normalised in order to match the count rates of each camera outside the target region and in the hard energy band of $[10-12] \si{\kilo\electronvolt}$. The normalization of each astrophysical component was determined using a joint spectral fit outside any cluster emission, in the same region where the particle background was normalised.

From the cleaned data we produced an image in the $[0.5-2.5]\si{\kilo\electronvolt}$, which is background subtracted and corrected for any spatial variation of the effective area.
From this image we determined the X-ray peak at ($RA=350.56228$, $DEC=28.52041$).
 Fig.~\ref{fig:xmm} shows a wavelet denoising of this image. As we can see from this Figure, the X-ray peak (circle) lies within 5$\arcsec$ of the galaxy that we assumed to be the BCG (cross), whereas the second brightest galaxy is 18$\arcsec$ away. It is worth noting that  there is also a good agreement between the direction of the elongation of the cluster, going from NE to the SW, as seen by the X-ray image and the isocontours from the lensing analysis. 

From the photon image we then extracted a point-source excised surface brightness profile, centered on the cluster X-ray peak.
 In Fig. ~\ref{fig:xprops}, top left panel, we show this surface brightness profile, which we found to reach the background level at $R=\SI{1450}{\kilo\parsec}$. The temperature profile, shown in Fig.~\ref{fig:xprops} bottom left panel, was extracted in 4 bins from the same
circular region as the surface brightness The temperature was measured in each bin by fitting the spectrum in the energy band $[0.3-11]\si{\kilo\electronvolt}$  using an absorbed APEC thermal model with three parameters, namely the normalization, temperature and metallicity. The thermal models were  convolved by the instrument response, as described in detail in \cite{mass_profiles}.
 The atomic hydrogen column density was fixed to its Galactic value $N_{H}=6.24 \times 10^{20} \si{\centi\meter}^{-2}$ (\citealt{lockman_1990}).

To measure the X-ray mass we followed the "forward" procedure described  in \cite{2010A&A...514A..93M} and \cite{rasia12}.  In short, analytic profiles of
the gas density and temperature \cite{vikh_prof} were projected using the spectroscopic like temperature \cite{mazzotta_wt} and fitted to the observed surface brightness and  projected temperature.   The  3D  information were then folded into the hydrostatic mass equation (e.g. \citealt{sarazin_1988} or \citealt{ettori_2013}). The confidence envelopes of the 3D profiles in Fig. \ref{fig:radmass} were computed following a Monte-Carlo procedure. More specifically, we generated several realisations of both a Poisson distributed surface brightness profile and a Gaussian distributed temperature profile, that hold the mean values and variances of the derived data quantities. For each of these realisations we fitted a new set of density and 3D temperature profiles, and stored these models when they obeyed the physical prior that the underlying hydrostatic mass profile is increasing and positive. From $N=500$ remaining hydrostatic density, 3D temperature and mass models, we drew the confidence envelopes that encompass $68\%$ of the realisations with minimal distance to their relative best fit. The envelope of the mass profile  allowed us to derive a cluster mass of $M_{2500}=3.2^{+0.8}_{-0.9} \times 10^{14}M_{\odot}$, $M_{1000}=4.4^{+0.8}_{-0.7} \times 10^{14}M_{\odot}$ and $M_{500}=5.4^{+1.0}_{-0.9} \times 10^{14}M_{\odot}$, which corresponds to $R_{2500}=541^{+49}_{-59}\si{\kilo\parsec}$, $R_{1000}=816^{+49}_{-46}\si{\kilo\parsec}$ and $R_{500}=1093^{+65}_{-62}\si{\kilo\parsec}$, respectively.
The latter mass value is in excellent agreement with the mass derived from iteration about the $M_{500}$ - $Y_X$ relation, $M_{\rm 500} = (5.60\pm0.22) \times 10^{14}$ $M_\odot$ \citep{planck2011-5.1b}.

 Figure~\ref{fig:radmass} also shows a comparison of the X--ray total mass profile  with those obtained from WL aperture mass densitometry and WL NFW model fitting (obtained with two free parameters).  
The inset shows the various mass estimates at density contrasts of 2500,
1000, and 500. The agreement between the different mass measurements is remarkably good at all density contrasts.

\citet{planck2013-p15} parameterised the difference between their mass estimate and the true mass by the factor (1-$b$), where $b$ contains all possible sources of bias, such as neglect of the non-thermal pressure support in the hydrostatic assumption, multi-temperature structure in the ICM, instrumental calibration uncertainties, etc. Assuming a  weak lensing reconstructed mass of $M_{\rm 500, WL} \sim (7\pm1) \times 10^{14}$ $M_\odot$ and comparing to the mass estimate obtained by \citet{planck2011-5.1b}, we obtain  (1-$b$) = 0.8 $\pm$ 0.1 from the present observations. 

\citet{planck2013-p15} obtained a bias factor of (1-$b$) = 0.8 from their calibration of $b$ using numerical simulations, and used this value, with a flat prior of $[0.7, 1.0]$, as the baseline for their cosmological analysis. \citet{planck2015-p24} have updated this analysis to include a calibration of $(1-b)$ using weak lensing mass measurements, taking into account sources of scatter such as triaxiality, the presence of subtructures, and systematic errors in the weak lensing measurement itself. Assuming the weak lensing mass to be the true mass, their analysis yields $(1-b)_{\rm WtG} = 0.69 \pm 0.07$ and  $(1-b)_{\rm CCCP} = 0.78 \pm 0.09$ for the samples of  \citet[][Weighing the Giants]{vdlinden14}  and \citet[][Canadian Cluster Cosmology Project]{hoe15}, respectively. Under the same assumption, the value we obtain from PLCKG100 is compatible within $1\sigma$ with the bias factor derived from these updated measurements.


\section{Summary and conclusions}
\label{sec:summary}

We presented here the first weak lensing analysis of the PLCK G100.2-30.4 cluster. 
The analysis confirmed that
this is a massive cluster ($M_{\rm vir} > 10^{15}$ $M_\odot$),  
in agreement with that expected from the richness estimate. 

At a density contrast of 500 with respect to the critical density of the Universe at the
cluster redshift, the weak lensing mass from aperture densitometry is
$M^{\rm ap}_{\rm500} = (6\pm2)  \times 10^{14}$ $M_\odot$, after deprojection. An NFW fit in the radial range 1-15 arcmin, shown by the
dotted line in Fig.~\ref{fig:radmass}, yields a mass  $M_{\rm 500} \sim (7\pm1) \times 10^{14}$ $M_\odot$. We checked the
robustness of the weak lensing mass estimates by repeating the analysis with  different choices of the cluster centre and of the NFW model fitting: in particular, we compared the results obtained by leaving both $c_{\rm vir}$ and $M_{\rm vir}$ as free parameters, and using different $(c, M, z)$ relations. These yield results consistent with those given above.

Comparing the mass derived by weak lensing with the value obtained by the reanalysis of XMM data, we obtain a bias factor of (1-$b$) = 0.8 $\pm$ 0.1. This is compatible within $1 \sigma$ with the bias factor obtained by \citet{planck2015-p24} from their calibration of (1-$b$) using newly-available weak lensing samples from \citet{vdlinden14} and \citet{hoe15}.  The present results illustrate the importance of high-quality data and the need for larger samples in such cluster mass studies.

\section {Acknowledgements}
We thank the referee for the helpful comments that improved the paper.
We acknowledge financial contributions from: contract ASI/INAF I/023/12/0 (LM, MR, MM); PRIN INAF 2009 "Weighing galaxy clusters with strong and weak lensing" (MR, MM);  PRIN MIUR 2010-2011 "The dark Universe and the cosmic evolution of baryons: from current surveys to Euclid" (LM, MR,MM); PRIN INAF 2012 "The Universe in the box: multiscale simulations of cosmic structure" (LM);
contracts ASI-INAF I/009/10/0 and PRIN-INAF 2012 "A unique dataset to address the most compelling open questions about X-Ray Galaxy Clusters" (SE). MR also acknowledges Prof. P. Benvenuti (Department of Physics and Astronomy, University of Padova), for having made available his computing facilities.

\bibliography{P100}

\begin{thebibliography}{69}
\expandafter\ifx\csname natexlab\endcsname\relax\def\natexlab#1{#1}\fi

\bibitem[{{Andreon} \& {Hurn}(2010)}]{2010MNRAS.404.1922A}
{Andreon}, S. \& {Hurn}, M.~A. 2010, \mnras, 404, 1922

\bibitem[{{Applegate} {et~al.}(2014){Applegate}, {von der Linden}, {Kelly},
  {Allen}, {Allen}, {Burchat}, {Burke}, {Ebeling}, {Mantz}, \&
  {Morris}}]{2014MNRAS.439...48A}
{Applegate}, D.~E., {von der Linden}, A., {Kelly}, P.~L., {et~al.} 2014,
  \mnras, 439, 48

\bibitem[{{Arnaud} {et~al.}(2010){Arnaud}, {Pratt}, {Piffaretti},
  {B{\"o}hringer}, {Croston}, \& {Pointecouteau}}]{2010A&A...517A..92A}
{Arnaud}, M., {Pratt}, G.~W., {Piffaretti}, R., {et~al.} 2010, \aap, 517, A92

\bibitem[{{Berg{\'e}} {et~al.}(2012){Berg{\'e}}, {Price}, {Amara}, \&
  {Rhodes}}]{2012MNRAS.419.2356B}
{Berg{\'e}}, J., {Price}, S., {Amara}, A., \& {Rhodes}, J. 2012, \mnras, 419,
  2356

\bibitem[{{Bertin}(2010)}]{SWarp}
{Bertin}, E. 2010, {SWarp: Resampling and Co-adding FITS Images Together},
  astrophysics Source Code Library

\bibitem[{{Bertin}(2011)}]{PSFex}
{Bertin}, E. 2011, in Astronomical Society of the Pacific Conference Series,
  Vol. 442, Astronomical Data Analysis Software and Systems XX, ed. I.~N.
  {Evans}, A.~{Accomazzi}, D.~J. {Mink}, \& A.~H. {Rots}, 435

\bibitem[{{Bertin} \& {Arnouts}(1996)}]{SExtractor}
{Bertin}, E. \& {Arnouts}, S. 1996, \aaps, 117, 393

\bibitem[{{Bourdin} \& {Mazzotta}(2008)}]{mass_profiles}
{Bourdin}, H. \& {Mazzotta}, P. 2008, \aap, 479, 307

\bibitem[{{Bourdin} {et~al.}(2013){Bourdin}, {Mazzotta}, {Markevitch},
  {Giacintucci}, \& {Brunetti}}]{bourdin_a521}
{Bourdin}, H., {Mazzotta}, P., {Markevitch}, M., {Giacintucci}, S., \&
  {Brunetti}, G. 2013, \apj, 764, 82

\bibitem[{{Broadhurst} {et~al.}(2008){Broadhurst}, {Umetsu}, {Medezinski},
  {Oguri}, \& {Rephaeli}}]{broad_2008}
{Broadhurst}, T., {Umetsu}, K., {Medezinski}, E., {Oguri}, M., \& {Rephaeli},
  Y. 2008, \apjl, 685, L9

\bibitem[{{Dickey} \& {Lockman}(1990)}]{lockman_1990}
{Dickey}, J.~M. \& {Lockman}, F.~J. 1990, \araa, 28, 215

\bibitem[{{Duffy} {et~al.}(2008){Duffy}, {Schaye}, {Kay}, \& {Dalla
  Vecchia}}]{2008MNRAS.390L..64D}
{Duffy}, A.~R., {Schaye}, J., {Kay}, S.~T., \& {Dalla Vecchia}, C. 2008,
  \mnras, 390, L64

\bibitem[{{Ettori} {et~al.}(2013){Ettori}, {Donnarumma}, {Pointecouteau},
  {Reiprich}, {Giodini}, {Lovisari}, \& {Schmidt}}]{ettori_2013}
{Ettori}, S., {Donnarumma}, A., {Pointecouteau}, E., {et~al.} 2013, \ssr, 177,
  119

\bibitem[{{Feldmann} {et~al.}(2006){Feldmann}, {Carollo}, {Porciani}, {Lilly},
  {Capak}, {Taniguchi}, {Le F{\`e}vre}, {Renzini}, {Scoville}, {Ajiki},
  {Aussel}, {Contini}, {McCracken}, {Mobasher}, {Murayama}, {Sanders},
  {Sasaki}, {Scarlata}, {Scodeggio}, {Shioya}, {Silverman}, {Takahashi},
  {Thompson}, \& {Zamorani}}]{Feldmann06}
{Feldmann}, R., {Carollo}, C.~M., {Porciani}, C., {et~al.} 2006, \mnras, 372,
  565

\bibitem[{{Formicola} {et~al.}(2014){Formicola}, Radovich, Meneghetti,
  Moscardini, Mazzotta, \& Grado}]{Formicola13}
{Formicola}, I., Radovich, M., Meneghetti, M., {et~al.} 2014, submitted

\bibitem[{{Giacconi} {et~al.}(2001){Giacconi}, {Rosati}, {Tozzi}, {Nonino},
  {Hasinger}, {Norman}, {Bergeron}, {Borgani}, {Gilli}, {Gilmozzi}, \&
  {Zheng}}]{giacconi}
{Giacconi}, R., {Rosati}, P., {Tozzi}, P., {et~al.} 2001, \apj, 551, 624

\bibitem[{{Girardi} {et~al.}(2012){Girardi}, {Barbieri}, {Groenewegen},
  {Marigo}, {Bressan}, {Rocha-Pinto}, {Santiago}, {Camargo}, \& {da
  Costa}}]{2012rgps.book..165G}
{Girardi}, L., {Barbieri}, M., {Groenewegen}, M.~A.~T., {et~al.} 2012,
  {TRILEGAL, a TRIdimensional modeL of thE GALaxy: Status and Future}, ed.
  A.~{Miglio}, J.~{Montalb{\'a}n}, \& A.~{Noels}, 165

\bibitem[{{Girardi} {et~al.}(2005){Girardi}, {Groenewegen}, {Hatziminaoglou},
  \& {da Costa}}]{2005A&A...436..895G}
{Girardi}, L., {Groenewegen}, M.~A.~T., {Hatziminaoglou}, E., \& {da Costa}, L.
  2005, \aap, 436, 895

\bibitem[{{Gruen} {et~al.}(2013){Gruen}, {Brimioulle}, {Seitz}, {Lee}, {Young},
  {Koppenhoefer}, {Eichner}, {Riffeser}, {Vikram}, {Weidinger}, \&
  {Zenteno}}]{Gruen13}
{Gruen}, D., {Brimioulle}, F., {Seitz}, S., {et~al.} 2013, \mnras, 432, 1455

\bibitem[{{Heymans} {et~al.}(2006){Heymans}, {Van Waerbeke}, {Bacon}, {Berge},
  {Bernstein}, {Bertin}, {Bridle}, {Brown}, {Clowe}, {Dahle}, {Erben}, {Gray},
  {Hetterscheidt}, {Hoekstra}, {Hudelot}, {Jarvis}, {Kuijken}, {Margoniner},
  {Massey}, {Mellier}, {Nakajima}, {Refregier}, {Rhodes}, {Schrabback}, \&
  {Wittman}}]{step1}
{Heymans}, C., {Van Waerbeke}, L., {Bacon}, D., {et~al.} 2006, \mnras, 368,
  1323

\bibitem[{{Hoekstra}(2003)}]{Hoekstra03}
{Hoekstra}, H. 2003, \mnras, 339, 1155

\bibitem[{{Hoekstra} {et~al.}(2000){Hoekstra}, {Franx}, \&
  {Kuijken}}]{apJ...532...88h}
{Hoekstra}, H., {Franx}, M., \& {Kuijken}, K. 2000, \apj, 532, 88

\bibitem[{Hoekstra {et~al.}(1998)Hoekstra, Franx, Kuijken, \&
  Squires}]{Hoekstra98}
Hoekstra, H., Franx, M., Kuijken, K., \& Squires, G. 1998, ApJ, 504, 636

\bibitem[{{Hoekstra} {et~al.}(2011){Hoekstra}, {Hartlap}, {Hilbert}, \& {van
  Uitert}}]{Hoekstra11}
{Hoekstra}, H., {Hartlap}, J., {Hilbert}, S., \& {van Uitert}, E. 2011, \mnras,
  412, 2095

\bibitem[{{Hoekstra} {et~al.}(2015){Hoekstra}, {Herbonnet}, {Muzzin}, {Babul},
  {Mahdavi}, {Viola}, \& {Cacciato}}]{hoe15}
{Hoekstra}, H., {Herbonnet}, R., {Muzzin}, A., {et~al.} 2015, \mnras, 449, 685

\bibitem[{{Hoekstra} {et~al.}(2012){Hoekstra}, {Mahdavi}, {Babul}, \&
  {Bildfell}}]{2012MNRAS.427.1298H}
{Hoekstra}, H., {Mahdavi}, A., {Babul}, A., \& {Bildfell}, C. 2012, \mnras,
  427, 1298

\bibitem[{{Huang} {et~al.}(2011){Huang}, {Radovich}, {Grado}, {Puddu},
  {Romano}, {Limatola}, \& {Fu}}]{Huang11}
{Huang}, Z., {Radovich}, M., {Grado}, A., {et~al.} 2011, \aap, 529, A93

\bibitem[{{Ilbert} {et~al.}(2009){Ilbert}, {Capak}, {Salvato}, {Aussel},
  {McCracken}, {Sanders}, {Scoville}, {Kartaltepe}, {Arnouts}, {Le Floc'h},
  {Mobasher}, {Taniguchi}, {Lamareille}, {Leauthaud}, {Sasaki}, {Thompson},
  {Zamojski}, {Zamorani}, {Bardelli}, {Bolzonella}, {Bongiorno}, {Brusa},
  {Caputi}, {Carollo}, {Contini}, {Cook}, {Coppa}, {Cucciati}, {de la Torre},
  {de Ravel}, {Franzetti}, {Garilli}, {Hasinger}, {Iovino}, {Kampczyk},
  {Kneib}, {Knobel}, {Kovac}, {Le Borgne}, {Le Brun}, {F{\`e}vre}, {Lilly},
  {Looper}, {Maier}, {Mainieri}, {Mellier}, {Mignoli}, {Murayama}, {Pell{\`o}},
  {Peng}, {P{\'e}rez-Montero}, {Renzini}, {Ricciardelli}, {Schiminovich},
  {Scodeggio}, {Shioya}, {Silverman}, {Surace}, {Tanaka}, {Tasca}, {Tresse},
  {Vergani}, \& {Zucca}}]{ilbert}
{Ilbert}, O., {Capak}, P., {Salvato}, M., {et~al.} 2009, \apj, 690, 1236

\bibitem[{{Ivezi{\'c}} {et~al.}(2007){Ivezi{\'c}}, {Smith}, {Miknaitis}, {Lin},
  {Tucker}, {Lupton}, {Gunn}, {Knapp}, {Strauss}, {Sesar}, {Doi}, {Tanaka},
  {Fukugita}, {Holtzman}, {Kent}, {Yanny}, {Schlegel}, {Finkbeiner},
  {Padmanabhan}, {Rockosi}, {Juri{\'c}}, {Bond}, {Lee}, {Stoughton}, {Jester},
  {Harris}, {Harding}, {Morrison}, {Brinkmann}, {Schneider}, \&
  {York}}]{2007AJ....134..973I}
{Ivezi{\'c}}, {\v Z}., {Smith}, J.~A., {Miknaitis}, G., {et~al.} 2007, \aj,
  134, 973

\bibitem[{Kaiser {et~al.}(1995)Kaiser, Squires, \& Broadhurst}]{kaiser95}
Kaiser, N., Squires, G., \& Broadhurst, T. 1995, ApJ, 449, 460

\bibitem[{{Kitching} {et~al.}(2012){Kitching}, {Balan}, {Bridle}, {Cantale},
  {Courbin}, {Eifler}, {Gentile}, {Gill}, {Harmeling}, {Heymans}, {Hirsch},
  {Honscheid}, {Kacprzak}, {Kirkby}, {Margala}, {Massey}, {Melchior},
  {Nurbaeva}, {Patton}, {Rhodes}, {Rowe}, {Taylor}, {Tewes}, {Viola},
  {Witherick}, {Voigt}, {Young}, \& {Zuntz}}]{2012MNRAS.423.3163K}
{Kitching}, T.~D., {Balan}, S.~T., {Bridle}, S., {et~al.} 2012, \mnras, 423,
  3163

\bibitem[{{Kitching} {et~al.}(2013){Kitching}, {Rowe}, {Gill}, {Heymans},
  {Massey}, {Witherick}, {Courbin}, {Georgatzis}, {Gentile}, {Gruen},
  {Kilbinger}, {Li}, {Mariglis}, {Meylan}, {Storkey}, \&
  {Xin}}]{2013ApJS..205...12K}
{Kitching}, T.~D., {Rowe}, B., {Gill}, M., {et~al.} 2013, \apjs, 205, 12

\bibitem[{{Kuntz} \& {Snowden}(2000)}]{kuntz_gala}
{Kuntz}, K.~D. \& {Snowden}, S.~L. 2000, \apj, 543, 195

\bibitem[{{Kuntz} \& {Snowden}(2008)}]{kuntz_particle}
{Kuntz}, K.~D. \& {Snowden}, S.~L. 2008, \aap, 478, 575

\bibitem[{{Leccardi} \& {Molendi}(2008)}]{xmm_model_lec_mol}
{Leccardi}, A. \& {Molendi}, S. 2008, \aap, 486, 359

\bibitem[{{Lumb} {et~al.}(2002){Lumb}, {Warwick}, {Page}, \& {De
  Luca}}]{lumb_power}
{Lumb}, D.~H., {Warwick}, R.~S., {Page}, M., \& {De Luca}, A. 2002, \aap, 389,
  93

\bibitem[{Luppino \& Kaiser(1997)}]{kaiser97}
Luppino, G. \& Kaiser, N. 1997, ApJ, 475, 20

\bibitem[{{Mahdavi} {et~al.}(2008){Mahdavi}, {Hoekstra}, {Babul}, \&
  {Henry}}]{2008MNRAS.384.1567M}
{Mahdavi}, A., {Hoekstra}, H., {Babul}, A., \& {Henry}, J.~P. 2008, \mnras,
  384, 1567

\bibitem[{{Mandelbaum} {et~al.}(2013){Mandelbaum}, {Rowe}, {Bosch}, {Chang},
  {Courbin}, {Gill}, {Jarvis}, {Kannawadi}, {Kacprzak}, {Lackner}, {Leauthaud},
  {Miyatake}, {Nakajima}, {Rhodes}, {Simet}, {Zuntz}, {Armstrong}, {Bridle},
  {Coupon}, {Dietrich}, {Gentile}, {Heymans}, {Jurling}, {Kent}, {Kirkby},
  {Margala}, {Massey}, {Melchior}, {Peterson}, {Roodman}, \&
  {Schrabback}}]{great10}
{Mandelbaum}, R., {Rowe}, B., {Bosch}, J., {et~al.} 2013, ArXiv e-prints

\bibitem[{{Mantz} {et~al.}(2010){Mantz}, {Allen}, {Rapetti}, \&
  {Ebeling}}]{2010MNRAS.406.1759M}
{Mantz}, A., {Allen}, S.~W., {Rapetti}, D., \& {Ebeling}, H. 2010, \mnras, 406,
  1759

\bibitem[{{Mazzotta} {et~al.}(2004){Mazzotta}, {Rasia}, {Moscardini}, \&
  {Tormen}}]{mazzotta_wt}
{Mazzotta}, P., {Rasia}, E., {Moscardini}, L., \& {Tormen}, G. 2004, \mnras,
  354, 10

\bibitem[{{Medezinski} {et~al.}(2010){Medezinski}, {Broadhurst}, {Umetsu},
  {Oguri}, {Rephaeli}, \& {Ben{\'{\i}}tez}}]{Medez10}
{Medezinski}, E., {Broadhurst}, T., {Umetsu}, K., {et~al.} 2010, \mnras, 405,
  257

\bibitem[{{Meneghetti} {et~al.}(2010){Meneghetti}, {Rasia}, {Merten},
  {Bellagamba}, {Ettori}, {Mazzotta}, {Dolag}, \&
  {Marri}}]{2010A&A...514A..93M}
{Meneghetti}, M., {Rasia}, E., {Merten}, J., {et~al.} 2010, \aap, 514, A93

\bibitem[{{Meneghetti} {et~al.}(2014){Meneghetti}, {Rasia}, {Vega}, {Merten},
  {Postman}, {Yepes}, {Sembolini}, {Donahue}, {Ettori}, {Umetsu}, {Balestra},
  {Bartelmann}, {Ben{\'{\i}}tez}, {Biviano}, {Bouwens}, {Bradley},
  {Broadhurst}, {Coe}, {Czakon}, {De Petris}, {Ford}, {Giocoli},
  {Gottl{\"o}ber}, {Grillo}, {Infante}, {Jouvel}, {Kelson}, {Koekemoer},
  {Lahav}, {Lemze}, {Medezinski}, {Melchior}, {Mercurio}, {Molino},
  {Moscardini}, {Monna}, {Moustakas}, {Moustakas}, {Nonino}, {Rhodes},
  {Rosati}, {Sayers}, {Seitz}, {Zheng}, \& {Zitrin}}]{2014ApJ...797...34M}
{Meneghetti}, M., {Rasia}, E., {Vega}, J., {et~al.} 2014, \apj, 797, 34

\bibitem[{{Miyazaki} {et~al.}(2002){Miyazaki}, {Komiyama}, {Sekiguchi},
  {Okamura}, {Doi}, {Furusawa}, {Hamabe}, {Imi}, {Kimura}, {Nakata}, {Okada},
  {Ouchi}, {Shimasaku}, {Yagi}, \& {Yasuda}}]{suprimecam}
{Miyazaki}, S., {Komiyama}, Y., {Sekiguchi}, M., {et~al.} 2002, \pasj, 54, 833

\bibitem[{{Okabe} {et~al.}(2010){Okabe}, {Takada}, {Umetsu}, {Futamase}, \&
  {Smith}}]{Okabe10}
{Okabe}, N., {Takada}, M., {Umetsu}, K., {Futamase}, T., \& {Smith}, G.~P.
  2010, \pasj, 62, 811

\bibitem[{{Ouchi} {et~al.}(2004){Ouchi}, {Shimasaku}, {Okamura}, {Furusawa},
  {Kashikawa}, {Ota}, {Doi}, {Hamabe}, {Kimura}, {Komiyama}, {Miyazaki},
  {Miyazaki}, {Nakata}, {Sekiguchi}, {Yagi}, \& {Yasuda}}]{2004ApJ...611..660O}
{Ouchi}, M., {Shimasaku}, K., {Okamura}, S., {et~al.} 2004, \apj, 611, 660

\bibitem[{{Planck Collaboration IX}(2011)}]{planck2011-5.1b}
{Planck Collaboration IX}. 2011, \aap, 536, A9

\bibitem[{{Planck Collaboration VIII}(2011)}]{planck2011-5.1a}
{Planck Collaboration VIII}. 2011, \aap, 536, A8

\bibitem[{{Planck Collaboration XX}(2014)}]{planck2013-p15}
{Planck Collaboration XX}. 2014, \aap, 571, A20

\bibitem[{{Planck Collaboration XXIV}(2015)}]{planck2015-p24}
{Planck Collaboration XXIV}. 2015, Submitted to \aap,
  [arXiv:astro-ph/1502.01597]

\bibitem[{Popesso {et~al.}(2005)Popesso, B\"oringher, Romaniello, \&
  Voges}]{AA.433.415P}
Popesso, P., B\"oringher, H., Romaniello, M., \& Voges, W. 2005, A\&A, 433, 415

\bibitem[{Radovich {et~al.}(2004)Radovich, Arnaboldi, Ripepi, Massarotti,
  McCracken, \& Mellier}]{virmosu}
Radovich, M., Arnaboldi, M., Ripepi, V., {et~al.} 2004, A\&A, 417, 51

\bibitem[{{Radovich} {et~al.}(2008){Radovich}, {Puddu}, {Romano}, {Grado}, \&
  {Getman}}]{Radovich08}
{Radovich}, M., {Puddu}, E., {Romano}, A., {Grado}, A., \& {Getman}, F. 2008,
  \aap, 487, 55

\bibitem[{{Rasia} {et~al.}(2012){Rasia}, {Meneghetti}, {Martino}, {Borgani},
  {Bonafede}, {Dolag}, {Ettori}, {Fabjan}, {Giocoli}, {Mazzotta}, {Merten},
  {Radovich}, \& {Tornatore}}]{rasia12}
{Rasia}, E., {Meneghetti}, M., {Martino}, R., {et~al.} 2012, New Journal of
  Physics, 14, 055018

\bibitem[{{Romano} {et~al.}(2010){Romano}, {Fu}, {Giordano}, {Maoli},
  {Martini}, {Radovich}, {Scaramella}, {Antonuccio-Delogu}, {Donnarumma},
  {Ettori}, {Kuijken}, {Meneghetti}, {Moscardini}, {Paulin-Henriksson},
  {Giallongo}, {Ragazzoni}, {Baruffolo}, {Dipaola}, {Diolaiti}, {Farinato},
  {Fontana}, {Gallozzi}, {Grazian}, {Hill}, {Pedichini}, {Speziali},
  {Smareglia}, \& {Testa}}]{Romano10}
{Romano}, A., {Fu}, L., {Giordano}, F., {et~al.} 2010, \aap, 514, A88+

\bibitem[{{Rowe} {et~al.}(2014){Rowe}, {Jarvis}, {Mandelbaum}, {Bernstein},
  {Bosch}, {Simet}, {Meyers}, {Kacprzak}, {Nakajima}, {Zuntz}, {Miyatake},
  {Dietrich}, {Armstrong}, {Melchior}, \& {Gill}}]{2014arXiv1407.7676R}
{Rowe}, B., {Jarvis}, M., {Mandelbaum}, R., {et~al.} 2014, ArXiv e-prints

\bibitem[{{Rozo} {et~al.}(2009){Rozo}, {Rykoff}, {Koester}, {McKay}, {Hao},
  {Evrard}, {Wechsler}, {Hansen}, {Sheldon}, {Johnston}, {Becker}, {Annis},
  {Bleem}, \& {Scranton}}]{2009ApJ...703..601R}
{Rozo}, E., {Rykoff}, E.~S., {Koester}, B.~P., {et~al.} 2009, \apj, 703, 601

\bibitem[{{Sarazin}(1988)}]{sarazin_1988}
{Sarazin}, C.~L. 1988, {X-ray emission from clusters of galaxies}

\bibitem[{Schlegel {et~al.}(1998)Schlegel, Finkbeiner, \& Davis}]{schlegel}
Schlegel, D., Finkbeiner, D., \& Davis, M. 1998, ApJ, 500, 525

\bibitem[{{Smith} {et~al.}(2001){Smith}, {Brickhouse}, {Liedahl}, \&
  {Raymond}}]{apec_code}
{Smith}, R.~K., {Brickhouse}, N.~S., {Liedahl}, D.~A., \& {Raymond}, J.~C.
  2001, \apjl, 556, L91

\bibitem[{{Snowden} {et~al.}(1995){Snowden}, {Freyberg}, {Plucinsky},
  {Schmitt}, {Truemper}, {Voges}, {Edgar}, {McCammon}, \&
  {Sanders}}]{snowden_gala}
{Snowden}, S.~L., {Freyberg}, M.~J., {Plucinsky}, P.~P., {et~al.} 1995, \apj,
  454, 643

\bibitem[{{Umetsu} {et~al.}(2014){Umetsu}, {Medezinski}, {Nonino}, {Merten},
  {Postman}, {Meneghetti}, {Donahue}, {Czakon}, {Molino}, {Seitz}, {Gruen},
  {Lemze}, {Balestra}, {Ben{\'{\i}}tez}, {Biviano}, {Broadhurst}, {Ford},
  {Grillo}, {Koekemoer}, {Melchior}, {Mercurio}, {Moustakas}, {Rosati}, \&
  {Zitrin}}]{2014ApJ...795..163U}
{Umetsu}, K., {Medezinski}, E., {Nonino}, M., {et~al.} 2014, \apj, 795, 163

\bibitem[{{Vikhlinin} {et~al.}(2006){Vikhlinin}, {Kravtsov}, {Forman}, {Jones},
  {Markevitch}, {Murray}, \& {Van Speybroeck}}]{vikh_prof}
{Vikhlinin}, A., {Kravtsov}, A., {Forman}, W., {et~al.} 2006, \apj, 640, 691

\bibitem[{{Vikhlinin} {et~al.}(2009){Vikhlinin}, {Kravtsov}, {Burenin},
  {Ebeling}, {Forman}, {Hornstrup}, {Jones}, {Murray}, {Nagai}, {Quintana}, \&
  {Voevodkin}}]{2009ApJ...692.1060V}
{Vikhlinin}, A., {Kravtsov}, A.~V., {Burenin}, R.~A., {et~al.} 2009, \apj, 692,
  1060

\bibitem[{{von der Linden} {et~al.}(2014){von der Linden}, {Mantz}, {Allen},
  {Applegate}, {Kelly}, {Morris}, {Wright}, {Allen}, {Burchat}, {Burke},
  {Donovan}, \& {Ebeling}}]{vdlinden14}
{von der Linden}, A., {Mantz}, A., {Allen}, S.~W., {et~al.} 2014, \mnras, 443,
  1973

\bibitem[{Wright \& Brainerd(2000)}]{Wright00}
Wright, C.~O. \& Brainerd, T.~G. 2000, ApJ, 534, 34

\bibitem[{{Yagi} {et~al.}(2002){Yagi}, {Kashikawa}, {Sekiguchi}, {Doi},
  {Yasuda}, {Shimasaku}, \& {Okamura}}]{2002AJ....123...66Y}
{Yagi}, M., {Kashikawa}, N., {Sekiguchi}, M., {et~al.} 2002, \aj, 123, 66

\bibitem[{{Zhang} {et~al.}(2010){Zhang}, {Okabe}, {Finoguenov}, {Smith},
  {Piffaretti}, {Valdarnini}, {Babul}, {Evrard}, {Mazzotta}, {Sanderson}, \&
  {Marrone}}]{2010ApJ...711.1033Z}
{Zhang}, Y.-Y., {Okabe}, N., {Finoguenov}, A., {et~al.} 2010, \apj, 711, 1033

\end{thebibliography}

\end{document}